\PassOptionsToPackage{pdfpagelabels=false}{hyperref} 
\documentclass{aa}

\usepackage[normalem]{ulem}
\usepackage{xcolor}
\usepackage{graphicx}
\usepackage{txfonts}
\usepackage{amsmath}
\usepackage{amssymb}	% Extra maths symbols
\usepackage{rotating}
\usepackage{natbib}
\usepackage{gensymb} 
\usepackage{textcomp}
\usepackage{multicol}

\bibpunct{(}{)}{;}{a}{}{,} % to follow the A&A style
\usepackage[colorlinks,linkcolor=blue,anchorcolor=blue,citecolor=blue]{hyperref}

\graphicspath{{./}{Figures}}

\begin{document}

% \label{firstpage}
% \pagerange{\pageref{firstpage}--\pageref{lastpage}}

\title{Non-thermal emission in galaxy groups at extremely low frequency: the case of A1213}

\author{T. Pasini\inst{\ref{inst1}}
\and V. H. Mahatma\inst{\ref{inst2}}
\and M. Brienza\inst{\ref{inst1}}
\and K. Kolokythas\inst{\ref{inst3}, \ref{inst4}}
\and D. Eckert\inst{\ref{inst5}}
\and F. de Gasperin\inst{\ref{inst1}} 
\and R. J. van Weeren\inst{\ref{inst6}}
\and F. Gastaldello\inst{\ref{inst7}}
\and D. Hoang\inst{\ref{inst8}}
\and R. Santra\inst{\ref{inst9}}
}
\authorrunning{T. Pasini, V. Mahatma, M. Brienza et al.}

%\author[0000-0003-2754-9258]{M.~Gaspari}

% List of institutions
\institute{ INAF - Istituto di Radioastronomia, via P. Gobetti 101, 40129, Bologna, Italy\label{inst1}\\ \email{thomas.pasini@inaf.it}
\and Cavendish Laboratory - Astrophysics Group, University of Cambridge, 19 JJ Thomson Avenue, Cambridge, CB3 0HE, United Kingdom\label{inst2}
\and Centre for Radio Astronomy Techniques and Technologies, Department of Physics and Electronics, Rhodes University,\\ P.O. Box 94, Makhanda 6140, South Africa\label{inst3}
\and South African Radio Astronomy Observatory, Black River Park North, 2 Fir St, Cape Town, 7925, South Africa\label{inst4}
\and Department of Astronomy, University of Geneva, Ch. d’Ecogia 16, 1290 Versoix, Switzerland\label{inst5}
\and Leiden Observatory, Leiden University, PO Box 9513, 2300 RA Leiden, The Netherlands\label{inst6}
\and INAF, IASF-Milano, via A.Corti 12, 20133 Milano, Italy\label{inst7}
\and Hamburger Sternwarte, Universität Hamburg, Gojenbergsweg 112, 21029 Hamburg, Germany\label{inst8}
\and National Centre for Radio Astrophysics, Tata Institute of Fundamental Research, Pune 411007, India\label{inst9}
}

%\date{Accepted 2020 July 8. Received 2020 July 8; in original form 2020 May 6}

%\pubyear{2022}

\abstract
{Galaxy clusters and groups are the last link in the chain of hierarchical structure formation. Their environments can be significantly affected by outbursts from AGN, especially in groups where the medium density is lower and the gravitational potential shallower. The interaction between AGN and group weather can therefore greatly affect their evolution.}
{We investigate the non-thermal radio emission in Abell 1213, a galaxy group which is part of a larger sample of $\sim$50 systems (X-GAP) recently granted XMM-Newton observations.}
{We exploit proprietary LOFAR 54 MHz and uGMRT 380 MHz observations, complementing them with 144 MHz LOFAR survey and XMM-Newton archival data.}
{A1213 hosts a bright AGN associated with one of the central members, 4C\,29.41, which was previously optically identified as a dumb-bell galaxy. Observations at 144 MHz at a resolution of 0.3$''$ allow us to resolve the central radio galaxy. From this source, a $\sim 500$ kpc-long tail extends North-East. Our analysis suggests that the tail likely originated from a past outburst of 4C\,29.41, and its current state might be the result of the interaction with the surrounding environment. The plateau of the spectral index distribution in the Easternmost part of the tail suggests mild particle re-acceleration, that could have re-energised seed electrons from the past activity of the AGN. While we observe a spatial and physical correlation of the extended, central emission with the thermal plasma, which might hint at a mini-halo, current evidence cannot conclusively prove this.}
{A1213 is only the first group, among the X-GAP sample, that we are able to investigate through low-frequency radio observations. Its complex environment once again demonstrates the significant impact that the interplay between thermal and non-thermal processes can have on galaxy groups.}


\maketitle

%%%%%%%%%%%%%%%%%%%%%%%%%%%%%%%%%%%%%%%%%%%%%%%%%%%%%%%%%%%%%%%%%%%%%%%%%%%%%%

\section{Introduction}

\label{sec:intro}

In the past decades, it has become clear that jets from Active Galactic Nuclei (AGN) can strongly affect the surrounding medium on a vast ($\sim$pc to $\sim$Mpc) range of physical scales \citep[e.g.,][]{Clark_1997, Zanni_2005, Morganti_2013, Brienza_2023}. This is especially relevant in galaxy clusters and groups, where jets can excavate through the Intra-Cluster or Intra-Group Medium (ICM or IGrM) creating cavities \citep[e.g.,][]{Bruggen_2002, Birzan_2004, Rafferty_2006}, uplifting metal-rich gas \citep[e.g.,][]{Ettori_2013, Gastaldello_2021} and inducing turbulence and shocks \citep[e.g.,][]{Brunetti_2007}.

While at the centre of galaxy clusters we primarily observe Faranoff-Riley I (FRI) sources, likely because initially-relativistic jets collect material and slow down while travelling through the high-density ICM, poorer environments such as those of galaxy groups often host FRII radio galaxies \citep{Yates_1989, Hill_1991}. Nevertheless, it is not rare to also find FRII in rich clusters (see e.g. \citealt{Owen_1995}), as evidence seems to show that the FRI-FRII dichotomy might also be linked to the SuperMassive Black Hole (SMBH) accretion mode \citep{Ghisellini_2001, Marchesini_2004}. Jetted AGN are indeed observed even in massive galaxies, which are predominantly hosted in richer environments \citep{Brienza_2023}. Feedback from these AGN jets plays therefore a fundamental role in the evolution of clusters and groups, promoting or preventing star formation and regulating the cooling of the ICM/IGrM (see e.g. \citealt{McNamara_2012} for a review).

Nevertheless, the non-thermal emission observed in clusters and groups is not only associated to radio galaxies (see e.g. reviews by \citealt{Feretti_2012, vanWeeren_2019}). Low-frequency instruments such as the Giant Metrewave Radio Telescope (GMRT, \citealt{Gupta_2017}) and the LOw-Frequency ARray (LOFAR, \citealt{vanHaarlem_2013}) have helped to shed light on diffuse, $\sim$Mpc-scale synchrotron emission that was originally detected by higher-frequency surveys \citep{Rengelink_1997, Giovannini_1999}. An example are radio halos, whose synchrotron spectra\footnote{$S_\nu \propto \nu^\alpha$, with $S_\nu$ being the flux density at frequency $\nu$ and $\alpha$ being the spectral index.} usually show a rather uniform spectral index with $\alpha \sim -1.2$ (see e.g. review by \citealt{vanWeeren_2019}), although this value was mainly estimated from high-frequency detections in massive systems. 

A growing number of recent detections by LOFAR and uGMRT, which started to find halos with relatively steep spectra ($\alpha < -1.5$, see e.g. \citealt{Bonafede_2012, Wilber_2018, diGennaro_2021, Bruno_2021, Edler_2022, Pasini_2022a, Pasini_2024}), combined with gamma-ray constraints \citep[e.g.,][]{Ackermann_2010, Zandanel_2014, Osinga_2024}, points to a leptonic origin where seed electrons get re-accelerated by turbulence induced by major mergers \citep{Brunetti_2001, Cassano_2005, Brunetti_2011}. This is in line with early predictions about Cosmic-Ray (CR) protons energy budget \citep{Brunetti_2008, Brunetti_2011}.

Radio relics are instead observed at the periphery of merging systems and always show a certain degree of polarisation and elongated morphology \citep{vanWeeren_2010}, with their orientation being usually orthogonal to that of the merger axis. The spectral index distribution shows a gradient, with steeper values in the region closer to the cluster centre, and flatter indices on the outer side \citep[e.g.,][]{Bonafede_2012, deGasperin_2014}. For these reasons, the most promising formation models invoke re-acceleration of seed electrons by shocks \citep{Brunetti_2007, Kang_2017} which, similarly to turbulence, are generated by major mergers.

While the vast majority of studies of continuum radio emission have been carried out on massive galaxy clusters, the mass regime of galaxy groups ($M_{500} < 10^{14} M_{\odot}$) is still largely unexplored. In that direction, the Complete Local-Volume Groups Sample (CLoGS) studies have explored the 
multi-wavelength \citep{OSullivanetal17,Kolokythasetal18,Kolokythasetal19,OSullivanetal18b,Kolokythasetal22} properties of 53 nearby central Brightest Group Galaxies (BGGs) examining their energy output in an unbiased set of low-mass systems. 

The impact of powerful AGN outbursts in groups, which may extend beyond the virial radius, has a significant effect on their evolution. Any difference with respect to more massive systems should therefore be reflected even in the properties of central radio galaxies \citep{Pasini_2020, Pasini_2021, Pasini_2022b} and of the Intra-Group Medium (IGrM, see e.g. review by \citealt{Eckert_2021}).

In light of these issues, the XMM-Newton Group AGN Project (X-GAP, \citealt{Eckert_2024}) aims to quantify the impact of AGN feedback in a complete sample of 49 galaxy groups, which will be observed with XMM-Newton for a total of $\sim$852 ks\footnote{The exposure time per system differs.}. The same systems have (or will) also been observed by LOFAR at 54 \citep{deGasperin_2023} and 144 MHz \citep{Shimwell_2022}, and a smaller sub-sample will also have dedicated uGMRT observations.

\begin{figure*}[t!]
\centering
    \includegraphics[scale=0.27]{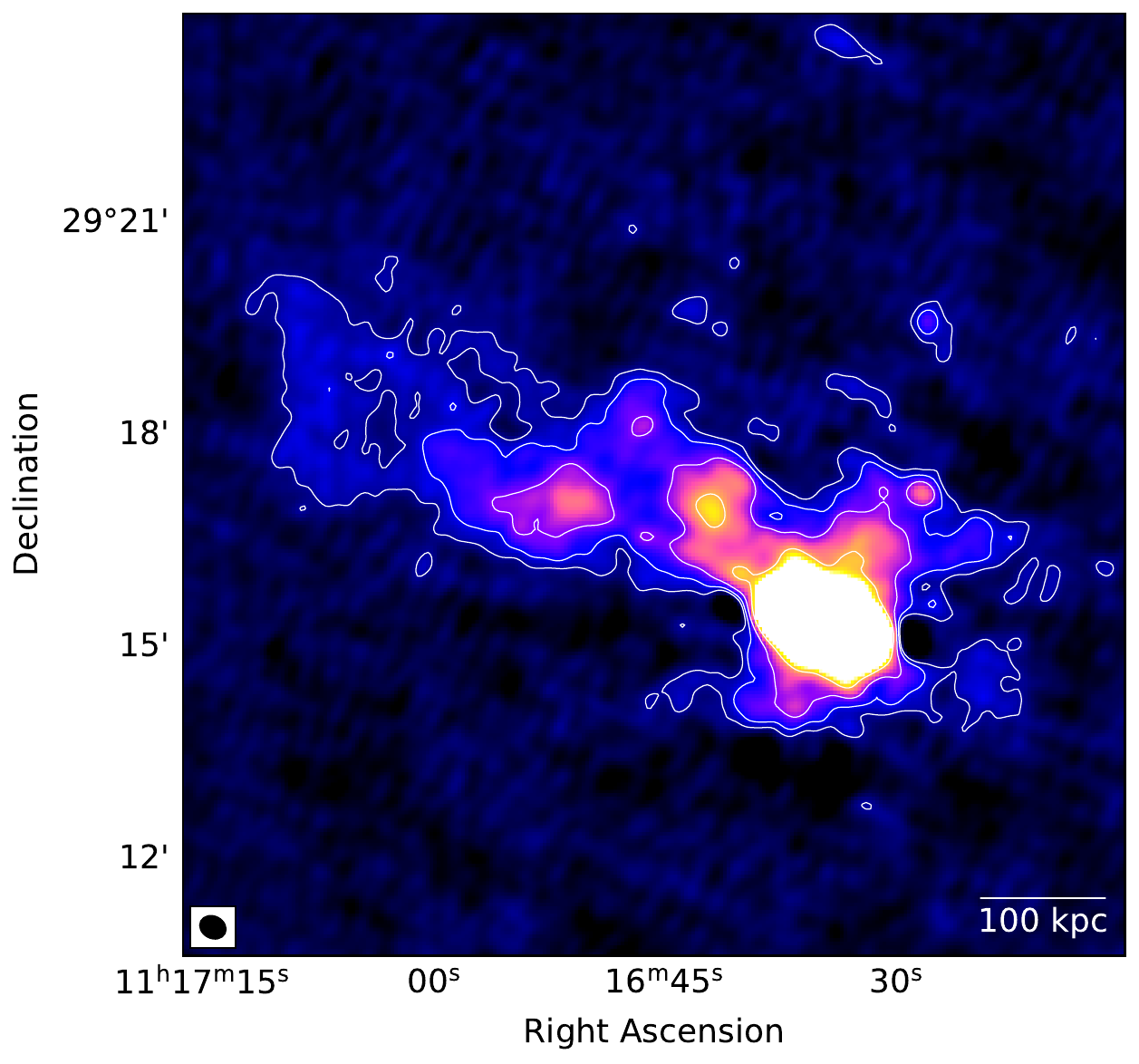}
    \includegraphics[scale=0.27]{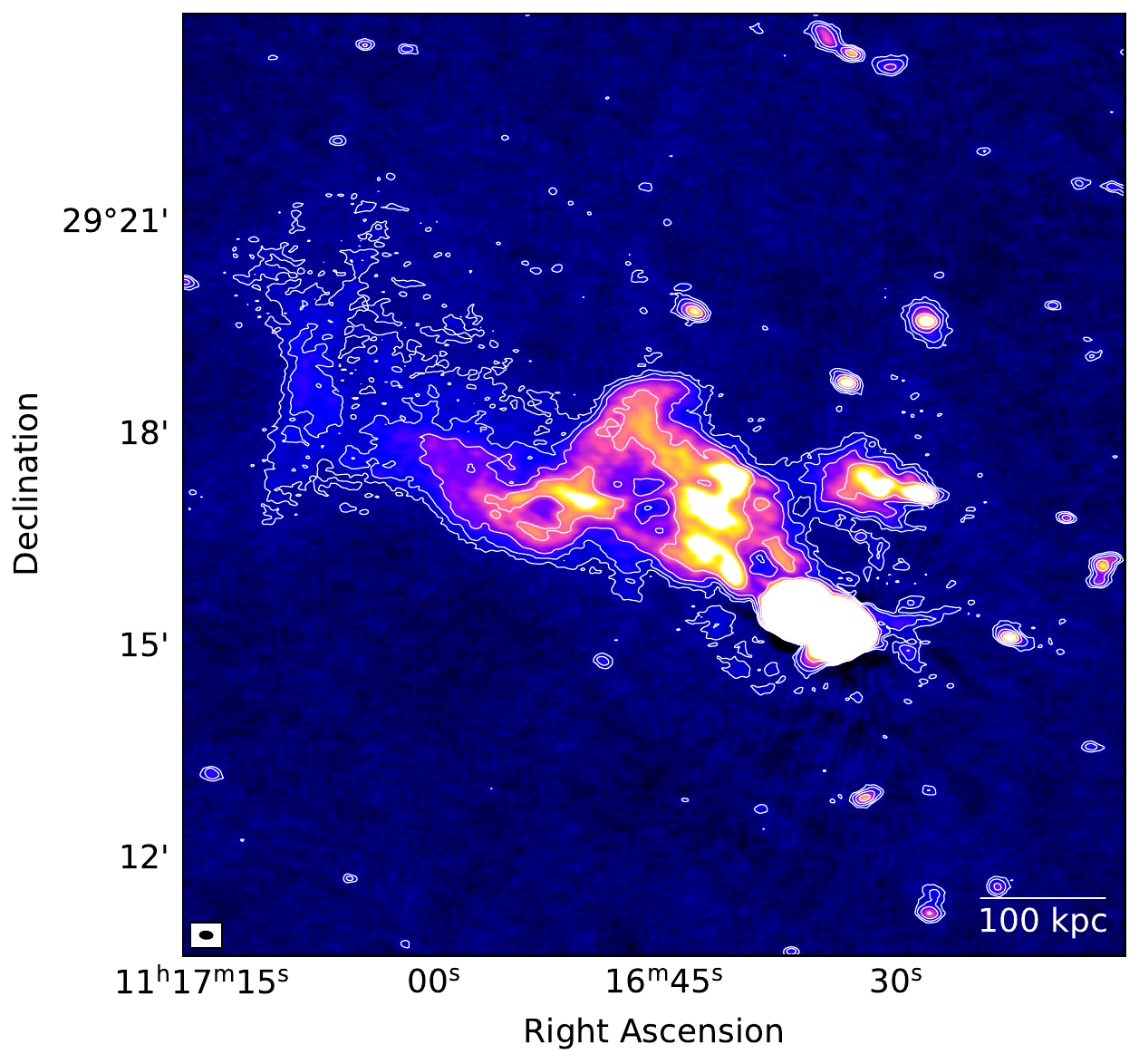}
    \includegraphics[scale=0.27]{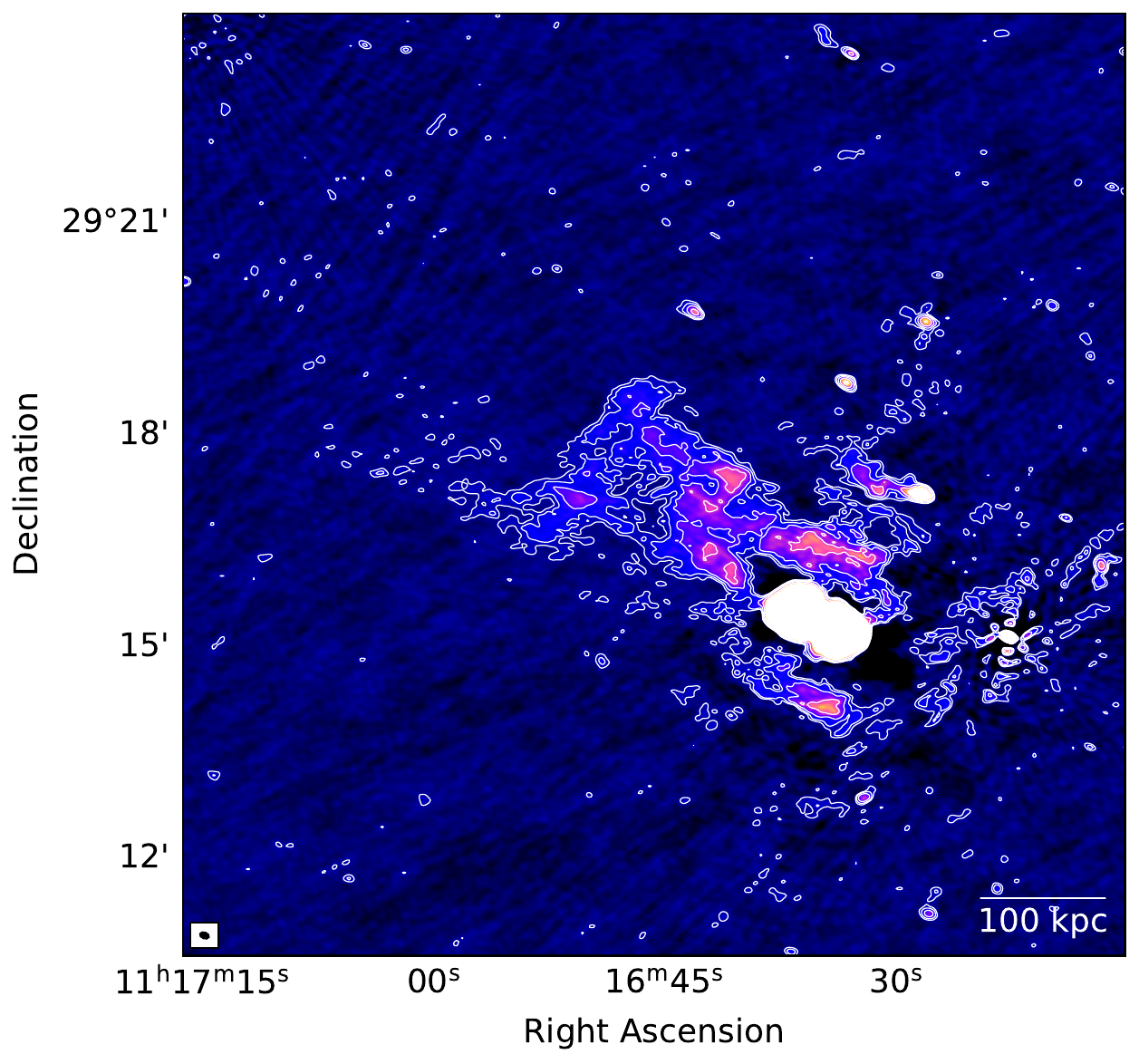}
    \caption{\textit{Left:} 54 MHz LOFAR image of A1213 produced with {\ttfamily briggs -0.3}. The \textit{rms} noise is $\sim$3 mJy beam$^{-1}$, and white contours are [-3, 3, 6, 12, 24...] $\times$ \textit{rms}. The beam is 22$''\times18''$. \textit{Central}: 144 MHz LOFAR image of A1213 produced with {\ttfamily briggs -0.3}. The \textit{rms} noise is $\sim$100 $\mu$Jy beam$^{-1}$, and white contours are [-3, 3, 6, 12, 24...] $\times$ \textit{rms}. The beam is 8$''\times5''$. \textit{Right:} 380 MHz uGMRT image of A1213 produced with {\ttfamily briggs -0.3}. The \textit{rms} noise is $\sim$60 $\mu$Jy beam$^{-1}$, and white contours are [-3, 3, 6, 12, 24...] $\times$ \textit{rms}. The beam is 7$''\times5''$.}
\label{fig:radio}
\end{figure*}

Abell 1213 (hereafter A1213) is a galaxy group\footnote{$R_{500} \sim 650$ kpc.} at \textit{z} = 0.0748 \citep{Abell_1989} which is part of the X-GAP sample. It lies on the edge between the cluster and group mass regimes ($M_{200} = (1.54 \pm 0.25) \times 10^{14} M_{\odot}$, \citealt{Boschin_2023}, hereafter B23). An early work by \citet{Giovannini_2009} first claimed the presence of diffuse emission, using 1.4 GHz observations at $\sim$35$''$ resolution by the Very Large Array (VLA)\footnote{in C and C/D configuration.}, to explain an elongated trail of emission which lies East of one of the brightest group members, 4C\,29.41 \citep{Jones-Forman_1999}, which they classify as a powerful FRII radio galaxy. The radio emission of this galaxy shows signs, in projection, of a physical association with the diffuse cluster plasma, which was labelled as a radio halo in \citet{Giovannini_2009}. The optical galaxy identified with the radio source 4C\,29.41 = B2 1113+29 is a dumb-bell galaxy \citep{Fanti_1982}. B23 discussed the presence of two nearby galaxies: ID 467 and ID 468. The radio emission is from ID 467. The two galaxies show an excess of intra-cluster light likely due to interaction and should be considered a couple of bright galaxies. On the other hand, they identified the BGG with an elliptical galaxy that lies $\sim$50 kpc West to 4C\,29.41. This object hosts a flat-spectrum radio source, although its morphology is rather compact ($\sim$15 kpc size) and looks disconnected from the more extended emission of 4C\,29.41.

B23 also studied the dynamics of this system, making use of optical data from the Sloan Digital Sky Survey (SDSS, \citealt{York_2000}). Thanks to a total of 143 spectroscopic galaxy members, they found that A1213 is a disturbed system consisting of several smaller galaxy groups, most likely merging together on the NE-SW direction. They support this hypothesis through an archival X-ray observation by XMM-Newton, which shows a clear elongated morphology of the IGrM in the same direction. Furthermore, through public 144 MHz LOFAR observations by the LOFAR Two-Metre Sky Survey \citep[LoTSS,][]{Shimwell_2019, Shimwell_2022}, they confirm the presence of large-scale ($\sim$500\,kpc) radio emission extending Eastwards from 4C\,29.41, but which does not show the typical properties and morphology of a radio halo. Its surface brightness distribution describes in fact an elongated and filamentary structure, that looks physically unrelated to the thermal plasma. B23 propose instead that this trail of emission is a radio relic. They support this hypothesis through a gradient in the spectral index map between 144 MHz and 1.4 GHz, which steepens from North to South. Finally, they interpret the presence of fragmented patches of emission around the BCG as a candidate, faint radio halo.

In this study, thanks to the availability of new low-frequency radio data, we propose a new interpretation on this very peculiar system, which is the first one within the X-GAP sample which we could study in detail thanks to LOFAR and uGMRT proprietary data. We perform a thorough analysis of the radio emission making use of a wide range of frequencies: from LOFAR Low Band Antenna (LBA) at 54 MHz, to LOFAR High Band Antenna (HBA) at 144 MHz, to uGMRT observations at 380 MHz. Thanks to the International Stations of LOFAR, we are able to reach an unprecedented angular resolution of $\sim$0.3$''$ at 144\,MHz for this system, which allows us to fully resolve the central radio galaxy.
Furthermore, we examine the properties of the large scale environment by performing a comparison between the non-thermal (radio) and the thermal (X-ray) emission using an archival XMM-Newton observation.

Throughout this work, we adopt a $\Lambda$CDM cosmology with H$_0 = 70$ km s$^{-1}$ Mpc$^{-1}$, $\Omega_\Lambda = 0.7$ and $\Omega_{\text{M}} =  1-\Omega_\Lambda  = 0.3$. Unless otherwise reported, errors are at 68$\%$ confidence level (1$\sigma$). At a redshift of \textit{z} = 0.0478 the luminosity distance is 212 Mpc, leading to a conversion of 1$''$ = 0.94 kpc.

%###################################################

\section{Data calibration and analysis}
\label{sec:calib}

\subsection{LBA observation}

A1213 has been observed by LOFAR in {\ttfamily LBA\_SPARSE\_EVEN} configuration on November 2$^{\rm nd}$, 2022, for a total of 10 hours as part of the proprietary project DDT18\_001 (P.I. H. Edler). The frequency range is $42 - 66$ MHz, with 54 MHz central frequency and 64 channels per sub-band, where each sub-band has a frequency width of 195 kHz.
The observation has been calibrated through the automated Pipeline for LOFAR LBA (PiLL), which is extensively described in \citet{deGasperin_2019, deGasperin_2021, deGasperin_2023}. We summarise here the main steps.

First, after a round of Radio Frequency Interferences (RFI) flagging and data averaging, bandpass and phase solutions for the calibrator (3C\,295) are estimated from a calibrator model included in PiLL, and then transferred to the target. Direction-independent calibration then corrects for first-order ionospheric delays, Faraday rotation and second-order beam errors. Direction-dependent calibration is performed by adopting the facet approach \citep{vanWeeren_2016}: all sources are subtracted from the visibilities, then the brightest source is re-added and solutions for this direction are determined. This step is repeated for every sufficiently bright source (DD-calibrator) in the field of view (FoV). The field is then divided into facets, based on the position of DD-calibrators, and the same solutions of each of them are applied to every source in the corresponding facet. These steps are done through {\ttfamily DDFacet} \citep{Tasse_2018}. 

\begin{figure*}[t!]
\centering
    \includegraphics[scale=0.38]{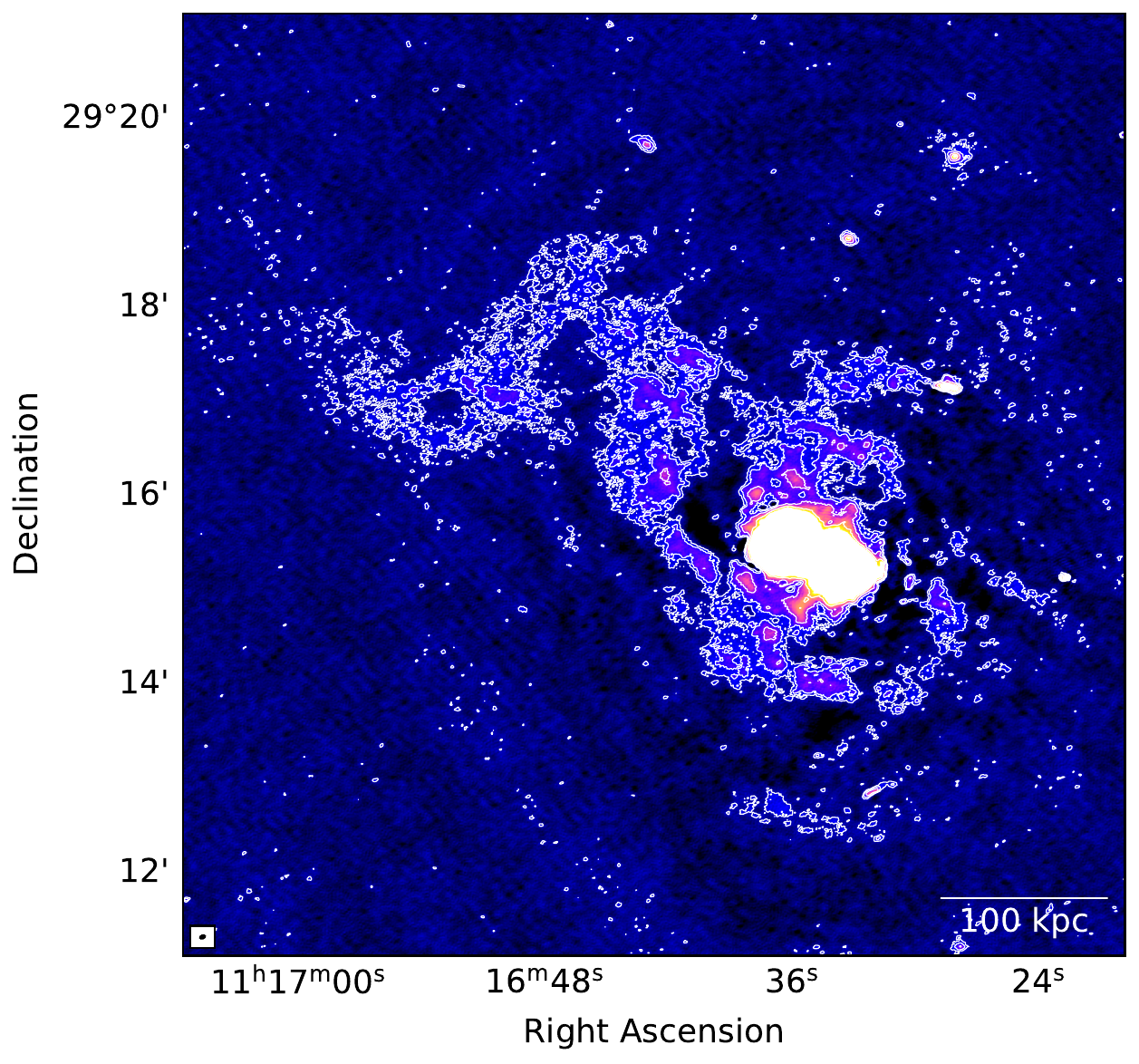}
    \hspace{0.3cm}
    \includegraphics[scale=0.38]{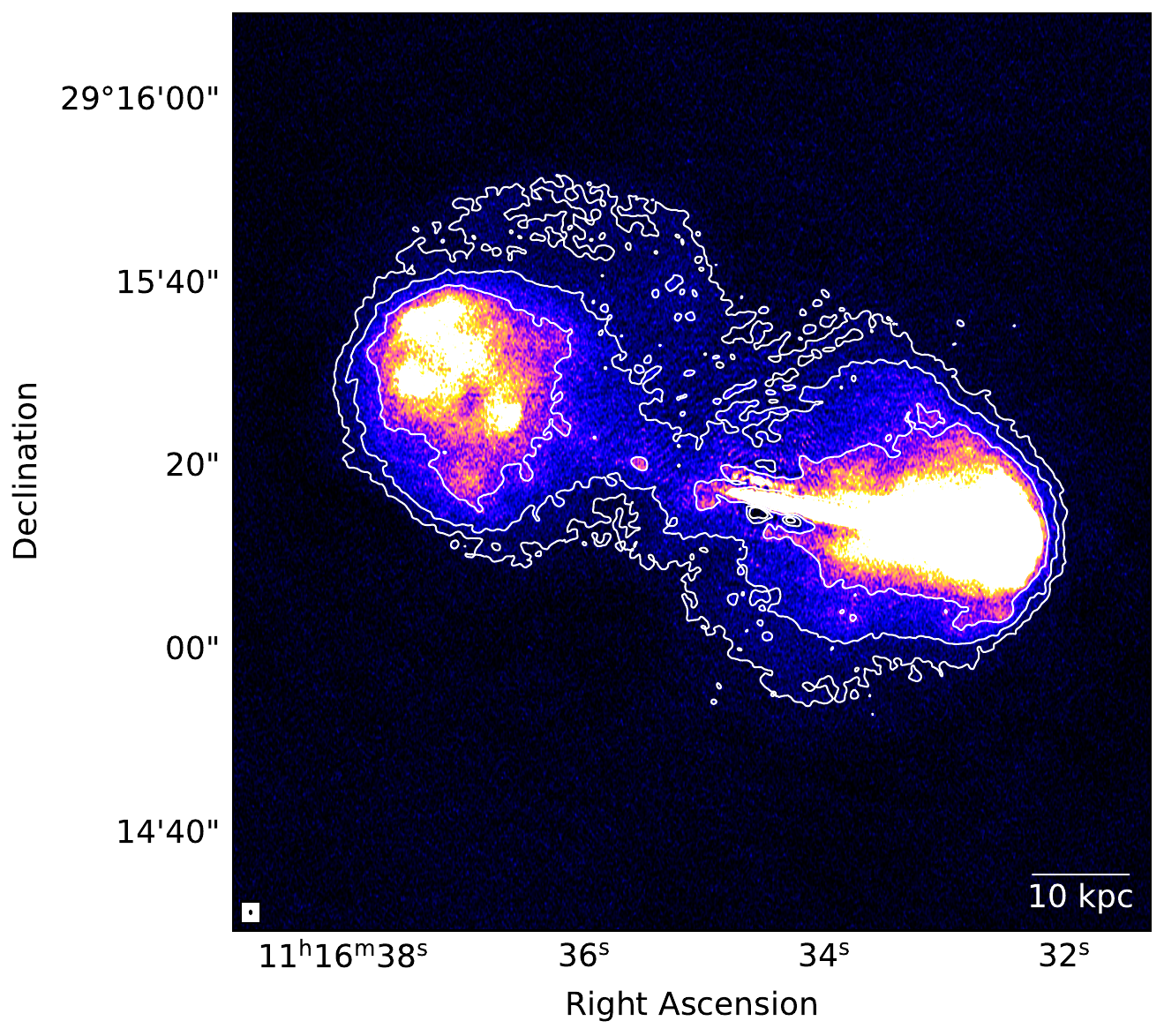}
    \caption{\textit{Left}: 144 MHz LOFAR-VLBI image of A1213. The image was produced with a restoring beam of 1.5$''\times1.5''$, and with an adapted colorscale to enhance the filamentary emission from the tail. Artefacts due to dynamic range issues are still visible in the image. The \textit{rms} noise is $\sim$50 $\mu$Jy beam$^{-1}$. \textit{Right}: 144\,MHz LOFAR-VLBI image of 4C\,29.41, the central radio galaxy of A1213. The image was produced at a resolution of 0.36$''\times0.19''$. The largest angular scale for this data is $\sim 70''$, which prevents the detection of the extended tail in the North-East. Contours are from 2$\sigma$ to highlight the presence of faint emission, with $\sigma \sim 45 \ \mu$Jy beam$^{-1}$ being the \textit{rms} noise.}
\label{fig:VLBI}
\end{figure*}

Finally, we further improve the correction of direction-dependent ionospheric effects by performing the same extraction process first adopted for HBA observations in \citet{vanWeeren_2021} and described for LBA in \citet{Pasini_2022a}. We select a small region ($\sim$ 15$'$) around our target of interest, A1213. All sources outside this region are subtracted and, if necessary, the phase centre of the observation is shifted to the target position. Data is averaged in time and frequency to reduce the overall size and smear the signal of improperly subtracted off-axis sources, and a primary beam correction at the new phase centre is applied through Image Domain Gridding (IDG, \citealt{vanderTol_2018}). Self-calibration cycles then help to reduce the noise and improve the image quality. 

Imaging is carried out through {\ttfamily WSClean} \citep{Offringa_2014} by applying suitable weighting and tapering of the visibilities in order to obtain images at different resolutions. To enhance the visualisation of extended diffuse emission, and avoid contamination by spurious sources, we performed the subtraction of compact sources as follows. First, an high-resolution image is produced by cutting visibilities below 800$\lambda$, which roughly corresponds to a Largest Angular Size (LAS) of 266$''$ and a Largest Linear Size of 250 kpc at the group redshift. This threshold was chosen, after visual inspection of the images, so that only sources below this LLS, which roughly corresponds to twice the size of the central AGN, are imaged. The clean components of these sources are then stored as a model through the {\ttfamily predict} option of {\ttfamily WSClean}, and subtracted from the original datasets. After this approach, images only show extended emission with LLS $>$ 250 kpc, while smaller sources get excluded. Flux density uncertainties are conservatively estimated to be 10\%, which is similar to LoLSS \citep{deGasperin_2021, deGasperin_2023}.

\subsection{HBA observation}
\label{sec:calibhba}

A1213 was observed by LOFAR at 144 MHz as part of the second Data Release of LoTSS DR2 \citep{Shimwell_2022}). The target is covered by two survey pointings, P168+30 and P171+30, for a total of 16 hours (8 hours per pointing, 8 hours longer than the observation used in B23 which only covers P168+30). The two pointings were independently calibrated making use of a set of automated pipelines developed to calibrate LOFAR HBA data: {\ttfamily prefactor} \citep{vanWeeren_2016, Williams_2016} and {\ttfamily ddf-pipeline} \citep{Tasse_2021}. The algorithm performs flagging of Radio Frequency Interferences (RFI), averages the data based on time and frequency, and correct for direction-independent effects such as clock offsets, complex gains and phase delays. The direction-dependent calibration follows the same steps listed above for LBA. Extraction is finally performed on the two separated datasets, similarly to what explained in the previous section, as well as imaging and compact-sources subtraction with the same threshold of 250 kpc used for the lower frequency. The flux density scale was aligned with the LoTSS-DR2 data release, where the flux calibration uncertainty is estimated to be $\sim$ 10\% \citep{Hardcastle_2021}. With respect to the image presented in B23, which is a simple cutout of the source in the pointing P168+30, our reprocessed image has better Signal to Noise (S/N), as it combines two survey pointings. For a comparison, the image in B23 has a \textit{rms} noise of $\sim$150 $\mu$Jy beam$^{-1}$, while we reach $\sim$100 $\mu$Jy beam$^{-1}$. Furthermore, the extraction has further improved the self-calibration of the source, eliminating most of the artefacts visible in the non-processed survey image.

The processing for the full European array of LOFAR is done using the LOFAR-VLBI pipeline\footnote{\url{https://github.com/LOFAR-VLBI/lofar-vlbi-pipeline}} described by \cite{Morabito_2022}, and deviates from the Dutch array processing described above only after the {\ttfamily ddf-pipeline} step. From the calibrated data which have the calibration solutions applied from {\ttfamily prefactor} and the {\ttfamily ddf-pipeline} (which are solutions for the Dutch stations only, except for Bandpass and ionospheric Rotation Measure), the phase centre of the data is then shifted to the coordinates of 4C\,29.41, and averaged. We then apply the \textit{facetselfcal} procedure \citep{vanWeeren_2021}, where we combine the core stations in the Dutch array into a `super station', allowing an accurate anchor for self-calibration of the longer baselines. This limits the largest angular scale to 71 arcsec. We begin a self-calibration cycle by using a starting sky model predicted from a VLA L-band image of 4C\,29.41, taken from the NVAS archive\footnote{\url{https://www.vla.nrao.edu/astro/nvas/}}. Several iterations of phase and complexgain self-calibration were performed, giving a final image with an RMS background noise of 50\,$\mu$Jy\,beam$^{-1}$ at an angular resolution of 0.36$''\times0.19''$.

To view the larger-scale cluster emission, we apply the beam corrections (array factor) and merged self-calibration solutions (as created above by \textit{facetselfcal}) determined in the direction of 4C\,29.41 and apply them to the data with {\ttfamily prefactor} and the {\ttfamily ddf-pipeline} (before combining Dutch core stations) solutions applied, in the manner described by \cite{Dejong_2024}. This corrected dataset is sensitive to angular scales up to $\gtrsim40$\,arcmin, since we did not combine the Dutch core stations. We then image with {\ttfamily WSClean} using a Gaussian tapering of 1.5\,arcsec, shown in the left panel of Figure \ref{fig:VLBI}. Note the cluster emission is resolved out at sub-arcsecond scales, prohibiting imaging and further self-calibration for residual errors.

\subsection{uGMRT observation}

uGMRT has observed A1213 on February 20$^{\rm th}$, 2022 for a total integration time of 4 hours in band 3 (central frequency $\sim$380 MHz), as part of the project 41\_093 (ObsID 13963, P.I. V. Mahatma). 3C\,147 was used as absolute flux density calibrator, and the data reduction and calibration were carried out using the Source Peeling and Atmospheric Modeling ({\ttfamily SPAM}) pipeline \citep{Intema_2009, Intema_2017}, which corrects for ionospheric effects and removes direction-dependent gain errors by adopting a facet approach, similarly to \citet{vanWeeren_2016}. Finally, data were corrected for the system temperature variations between calibrators and target, and imaging was carried out through WSClean applying different weightings and visibility tapering. Flux density uncertainties are assumed to be 6\%, as from literature values for uGMRT band 3 \citep[e.g.,][]{Chandra_2004, Bruno_2023}.

\subsection{XMM-Newton observation}

A1213 was observed by \emph{XMM-Newton} on June 12, 2008 for a total of 20ks (observation ID 0550270101). We processed the data using the \texttt{XMMSAS} v19.1 package and the X-COP analysis pipeline \citep{Ghirardini:2019}. After applying the standard cuts for time intervals affected by background flares, the clean observing time is 15.6 ks for the two MOS detectors and 5.3 ks for the PN detector. Our analysis procedure follows that of B23. From the cleaned event lists, we extracted count maps in the [0.7-1.2] keV band. For a discussion of the global IGrM properties of the system (e.g. hydrostatic mass, luminosity, surface brightness and density profiles), we refer to B23. In this work, the X-ray map has mainly been used for visualization purposes and for a comparison with the non-thermal emission detected in the radio band.

%###################################################

\section{Results}

\begin{figure*}
\centering
    \includegraphics[scale=0.403]{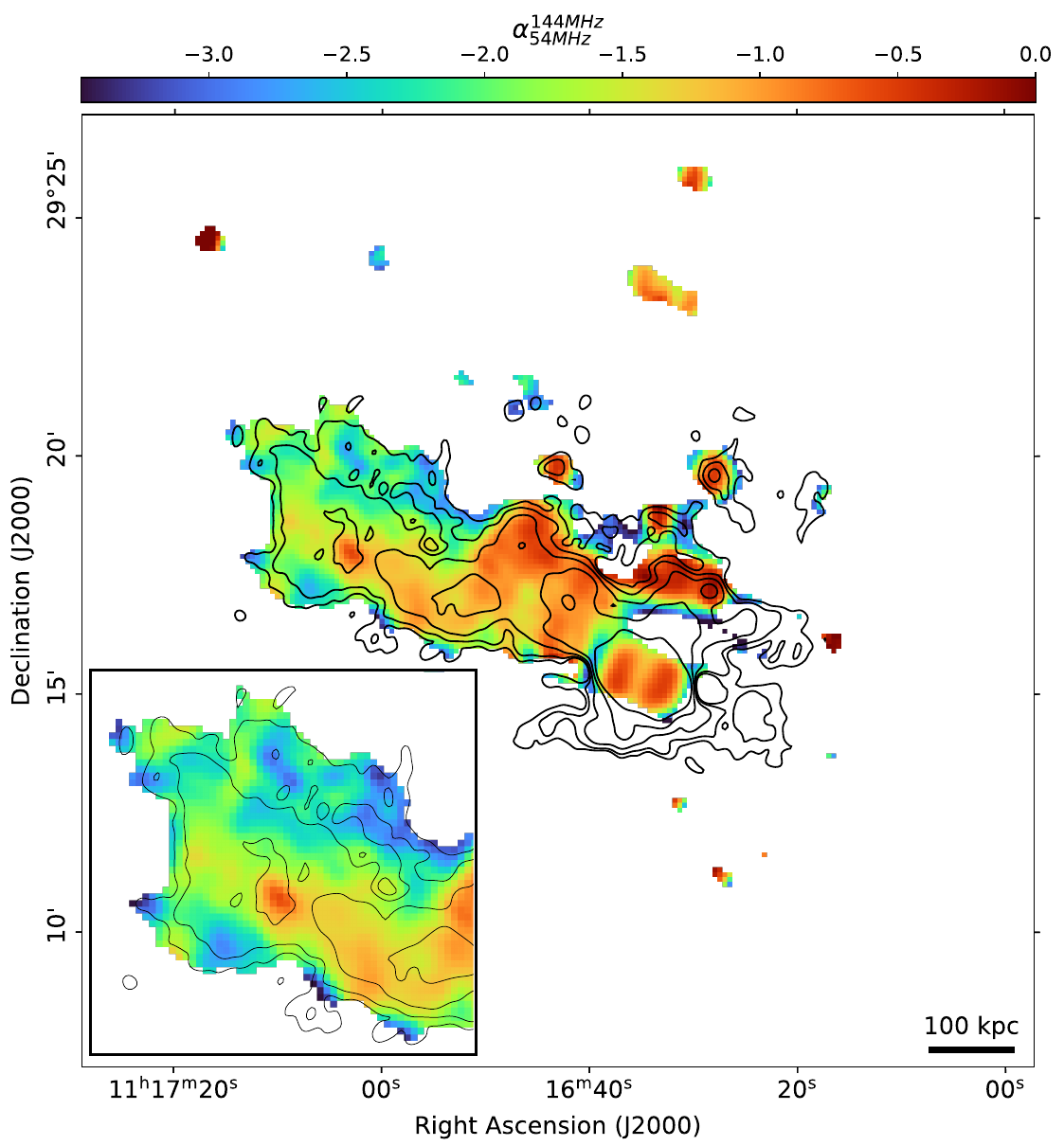}
    \hspace{1cm}
    \includegraphics[scale=0.35]{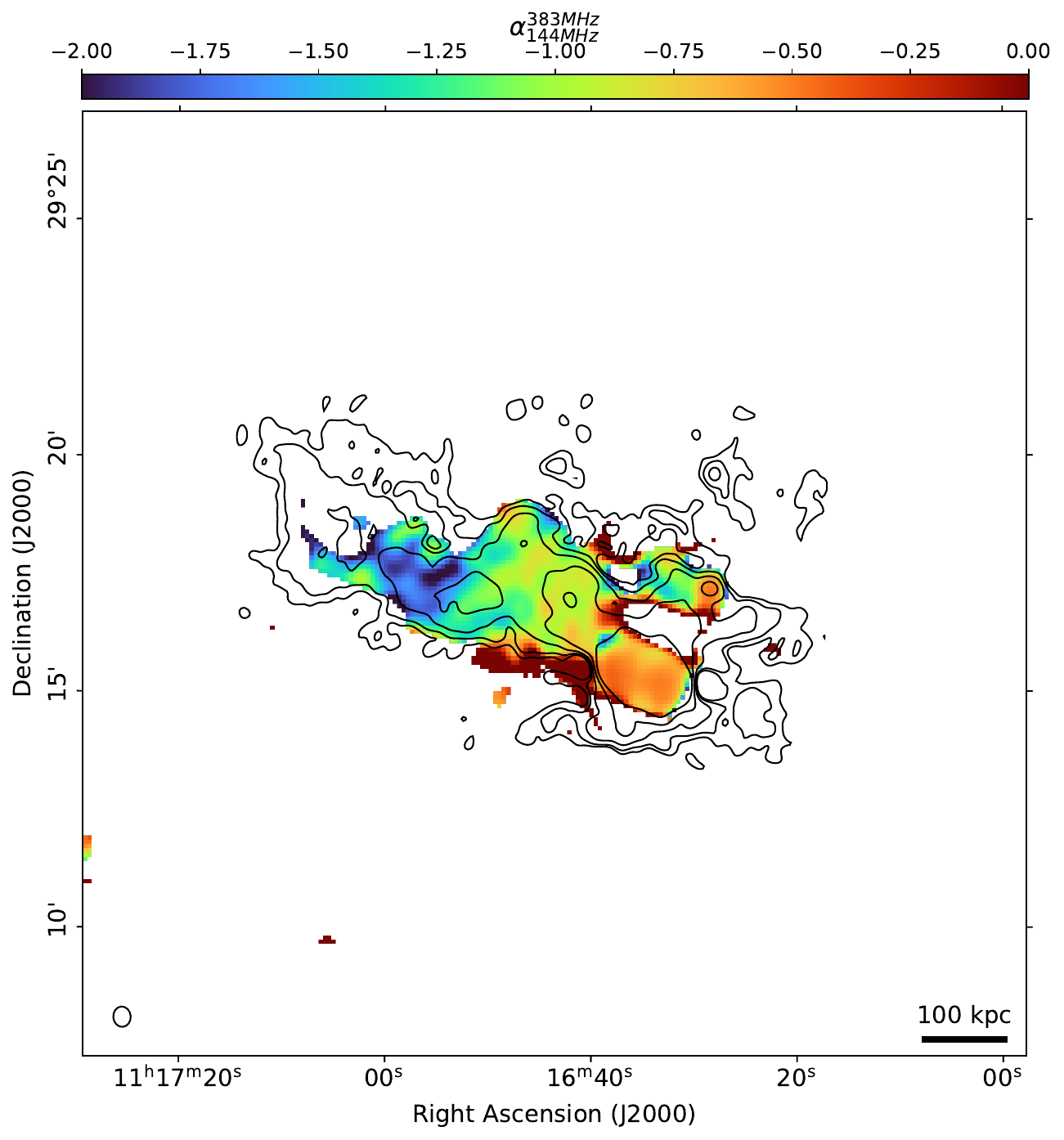}
    \caption{\textit{Left}: Spectral index map between 54 and 144 MHz (beam $30'' \times 25''$). Upper limits are represented with downwards arrows. Only emission above 3$\sigma$ is visible. Overlaid in black are 54 MHz contours. The bottom-left panel is a zoom-in on the Eastern part of the tail. \textit{Right}: Spectral index map between 144 and 380 MHz (beam $24'' \times 19''$). Only emission above 3$\sigma$ is visible. Overlaid in black are 54 MHz contours. In both images the emission North and South to 4C\,29.41, which was affected by calibration artefacts at 144 and 380 MHz, has been masked.}
\label{fig:spidx}
\end{figure*}

In this section we summarise the results of the analysis of the 54, 144 and 380 MHz images, as well as the spectral index maps we produced. The synchrotron spectrum of 4C\,29.41, given its peculiarity, will be discussed in detail in Sec. \ref{sec:synch}.

\subsection{A1213 at 54 MHz}

In the left panel of Fig. \ref{fig:radio} we show the 54 MHz LOFAR image of A1213.  The morphology of the tail is very similar to what is observed at higher frequency (see B23 and next sections), with a $\sim$500 kpc trail of emission extending North-East from the radio galaxy (at least in projection). Multiple peaks are visible along the tail. For this source we measure a flux density of $S_{54 \rm MHz}= 4.4 \pm 0.4$ Jy within 3$\sigma$ contours \footnote{excluding the central AGN.}. North and South to 4C\,29.41 we observe instead elliptical emission with orthogonal orientation with respect to the AGN and the tail, which might suggest a different nature. This emission is partially blended with a head-tail radio galaxy North-East of 4C\,29.41. Indeed, B23 had already observed this 'patched' emission at 144 MHz, although the image was not clear due to calibration artefacts (see also next section). Our calibration strategy for the 54\,MHz observation helped to mitigate this effect, which allows for a clearer detection. The detected emission is therefore real and not the result of bad calibration.

\begin{figure*}[t!]
\centering
    \includegraphics[scale=0.7]{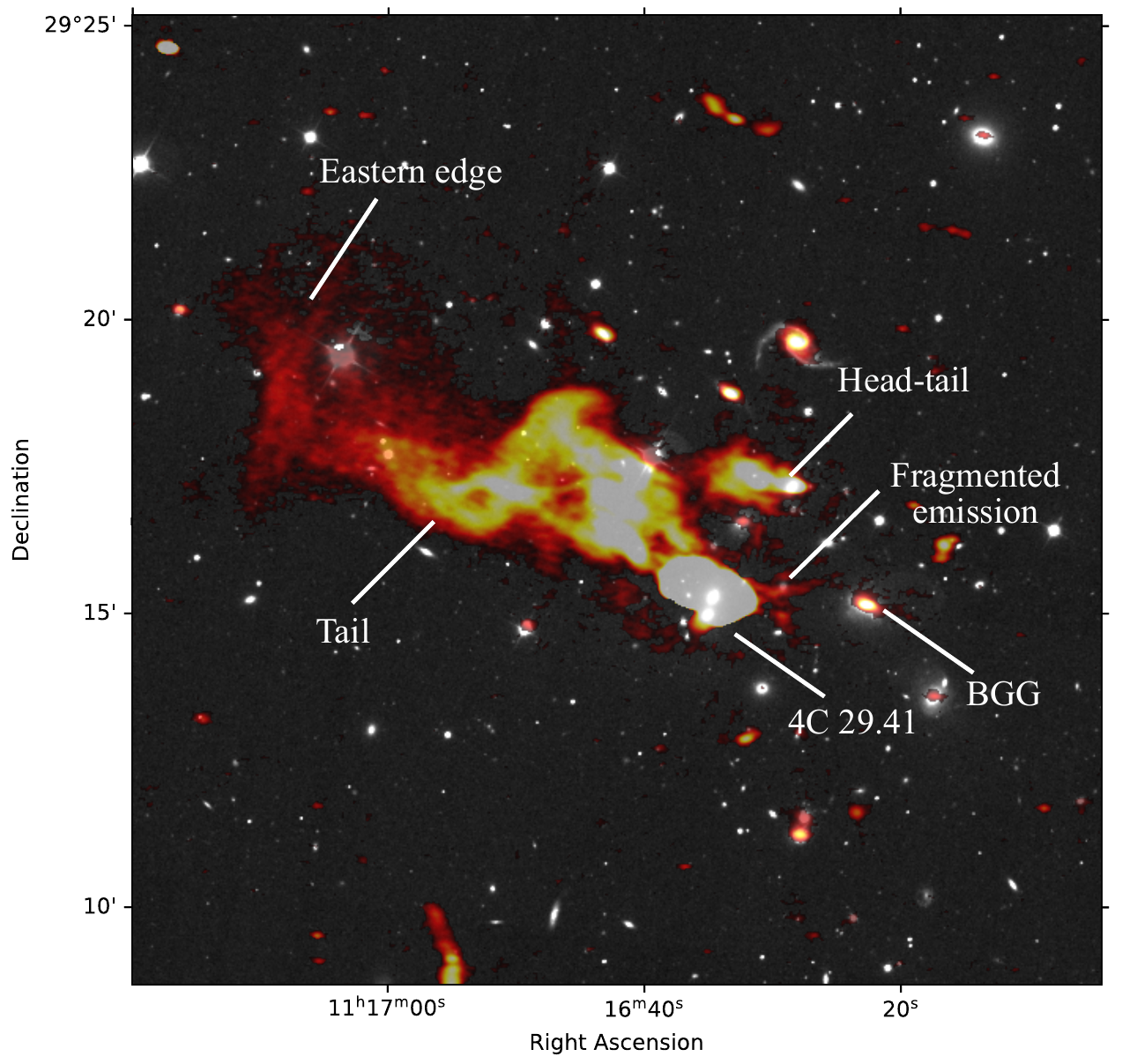}
    \caption{Radio emission from LOFAR at 144 MHz (red) overlaid on the optical image of A1213 from SDSS (\textit{g}-band). Relevant features that we discuss in this work have been labelled.}
\label{fig:large}
\end{figure*}

\subsection{A1213 at 144 MHz}
\label{sec:hbavlbi}

In the central panel of Fig. \ref{fig:radio} we show the 144 MHz map of A1213. Since a detailed discussion of the 144 MHz radio emission at at a resolution of $\sim 6''$ can be already found in B23, we summarise here the main features. The Eastern trail shows a number of emission peaks and a filamentary structure. Moreover, at the Eastern tip the radio emission seems to blend with a structure oriented in the N-S direction. This structure is also detected at 54 MHz, although at lower resolution. We confirm that the small-scale emission North to 4C\,29.41 is associated to a member galaxy (ID 442 in B23), and the detection of fragmented emission around the AGN, although that area is contaminated by calibration artefacts. We note that, even after our calibration strategy, we do not detect any new radio source in our image compared to B23, while we confirm the detection of all the sources they already found, even at lower and higher frequency. For the tail, and by excluding 4C\,29.41, we measure a flux density of $S_{144 \rm MHz} = 1.52 \pm 0.15$ Jy within 3$\sigma$ contours.

Thanks to the inclusion of LOFAR international stations, we are able to increase the resolution of the 144 MHz data. In the left panel of Fig. \ref{fig:VLBI} we show the LOFAR-VLBI image at 1.5$''$ resolution, which confirms the filamentary structure of the tail. It is also worth noting that we clearly detect emission North and South of 4C\,29.41, which is not visible at the typical 6$''$ resolution of the HBA image in the central panel of Fig. \ref{fig:radio}. Nevertheless, calibration artefacts are still visible in the image. In the right panel we show instead the LOFAR-VLBI image produced by phasing-up the core stations of the LOFAR Dutch array (see Sec. \ref{sec:calibhba} for details). While with this strategy the largest angular scale becomes smaller (i.e. $\sim 70''$), implying that we cannot detect the extended tail, we are able to further push the resolution to 0.3$''$, better calibrate the data and resolve the central radio galaxy. We observe a typical double-lobed, FRII-like morphology, with two jets (among which only one is visible) departing in the North East - South West direction for $\sim$35 kpc each. At this distance from the AGN core, the interaction with the surrounding IGrM creates two roughly spherical lobes with $\sim$40 kpc diameter. While the brightest radio emission is well-confined within the lobes, at 2$\sigma$ we detect fainter structures North and South of the East and West lobe, respectively. Finally, at this resolution the AGN does not look connected to the tail.

\subsection{A1213 at 380 MHz}

In the right panel of Fig. \ref{fig:radio} we show the uGMRT map of A1213 at 380 MHz. The tail shows a complex, filamentary structure which extends for $\sim$300 kpc (at 3$\sigma$), significantly less than what observed at lower frequency. This is just the result of the shape of the synchrotron spectrum, as higher energy electrons lose energy faster. We measure a flux density of $S_{380 \rm MHz} = 0.4 \pm 0.1$ Jy within 3$\sigma$ contours. We do not detect the N-S structure at the Eastern end of the tail, which might also suggest a steep spectrum (see also Sec. \ref{sec:relic}). We highlight the presence of emission North and South to 4C\,29.41 even at this frequency but, similarly to the HBA data, dynamic range issues affect the images and produce patches and negatives that prevent us from observing its real morphology.

\subsection{Spectral index maps}
\label{sec:spindex}

\begin{figure*}[t!]
\centering
    \includegraphics[scale=0.65]{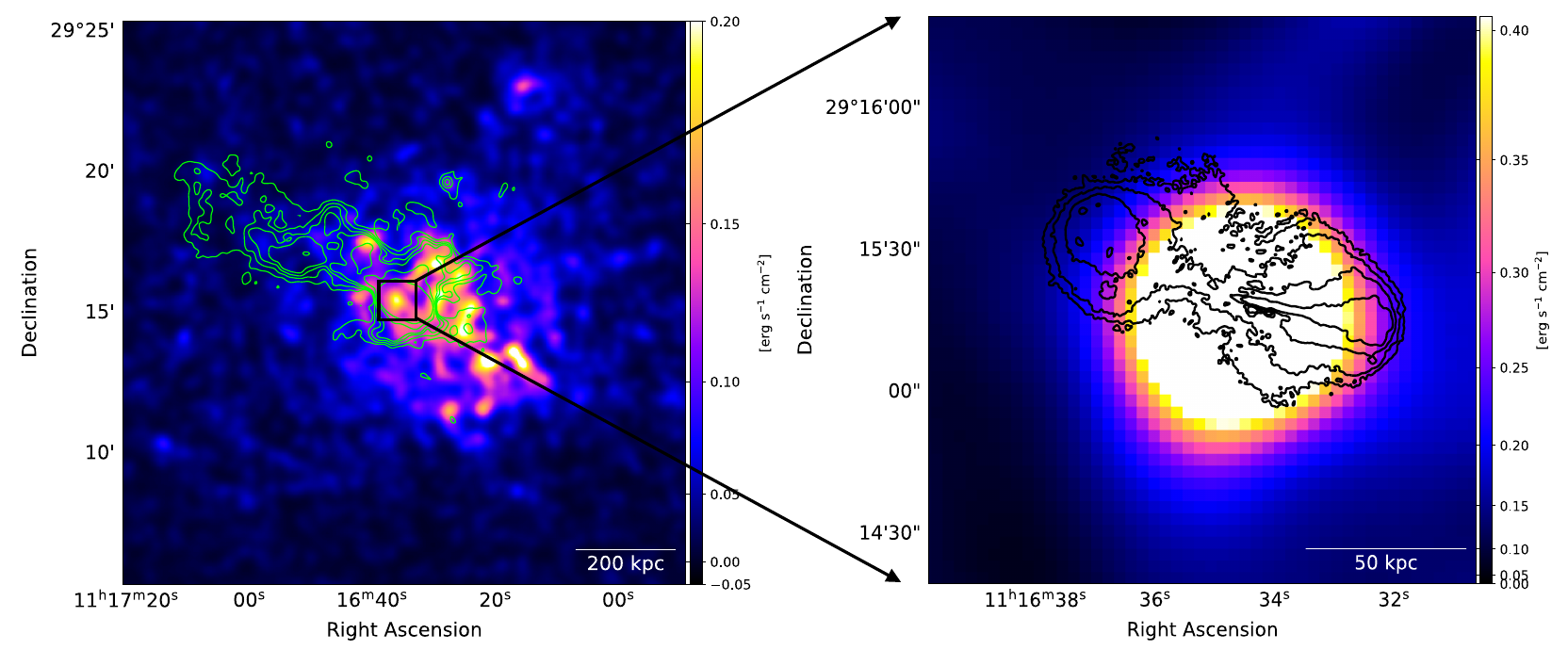}
    \caption{\textit{Left:} LOFAR 54 MHz contours (green) overlaid on the XMM-Newton map in the energy range 0.7-1.2 keV. \textit{R$_{200}$} is $\sim$1140 kpc, which lies outside of our field of view. \textit{Right:} zoom-in on 4C\,29.41 contours as detected at 144 MHz from the 0.3$''$ resolution image, overlaid on the same XMM-Newton map shown on the left panel}.
\label{fig:overlay}
\end{figure*}

We have produced spectral index maps between 54, 144 and 380 MHz as follows. First, maps at all frequencies were produced by applying the same visibility cut of 80$\lambda$-14k$\lambda$ and {\ttfamily Briggs -0.3}\footnote{Chosen to emphasize the extended emission.}. This is important in the comparison of images obtained with a different sensitivity to a diffuse low brightness emission, as the two instruments (LOFAR and uGMRT) have different \textit{uv}-coverages. The spectral index map was then generated by calculating $\alpha$ and $\Delta\alpha$ in each pixel. Additional details can be found in Appendix \ref{app:a}.

\subsubsection{LOFAR spectral index map between 54 and 144 MHz}

The spectral index map between 54 and 144 MHz is shown in the left panel of Fig. \ref{fig:spidx}, while the spectral index error map can be found in Appendix \ref{app:a}\footnote{Median uncertainties are $\Delta \alpha \sim 0.15$.}. It has been produced at the smallest common beam ($30'' \times 25 ''$) at which all sources of interest are clearly visible. 

We find that for 4C\,29.41\footnote{Radio galaxy only, excluding the tail.} the spectral index is $\sim$-0.8, consistent with what usually detected in active radio galaxies \citep{Zajacek_2019}. The spectral index then steepens moving Eastwards, and we detect what seems to be a sharp break in the gradient in a small region in-between 4C\,29.41 and the tail, with steep values of $\alpha \sim -2.3$. Interestingly, the Southern part of the tail steepens down to $\alpha \sim -1.2$ towards East, settling around this value. On the other hand, in the Northernmost part there is a further steepening from $\sim -2$ down to a minimum value of $\sim$ -3.1. Finally, the Eastern edge of the tail shows steep spectral indices with mean value $\alpha \sim -2$, with the exception of a thin filament oriented in the North-South direction which shows $\alpha \sim -1.2$. This corresponds to the structure previously detected at 144 MHz.

\subsubsection{LOFAR-uGMRT spectral index map between 144 and 380 MHz}

In the right panel of Fig. \ref{fig:spidx} we show the spectral index map between 144 and 380 MHz. This has been produced with the same method described for the LOFAR spectral index map, and by applying a visibility cut of 80 $\lambda$-14 k$\lambda$. We have masked the extended emission North and South to 4C\,29.41, since it is affected by dynamic range issues at both frequencies.

As expected, there is less spectral index information at higher frequencies due to the non-detection of steep-spectrum emission. Nevertheless, the spectral index trend is very similar to what is observed at lower frequency. The extended emission East of 4C\,29.41 shows values that steepen from $\sim$-0.7 down to minimum values of $\sim$-2 at the Eastern tip. However, at this higher frequency and distance from the AGN core ($\sim$100 kpc) the width of the tail looks smaller, and it is harder to detect any kind of gradient on the North-South direction, as instead observed with LOFAR, although the Northernmost region shows $\alpha \sim -1.4$. A significant gradient is instead detected on the West-East axis, i.e. moving from the AGN to the tip of the tail, with the spectral index gradually steepening, similarly to what observed at lower frequency.

The spectral index of 4C\,29.41 is consistent with what observed with LOFAR HBA and LBA. We note that we still detect, even at this frequency, a very small region in-between the AGN and the tail in which there is a break in the spectral index gradient, with values that suddenly steepen down to $\alpha \sim -2$. This region is as wide as the restoring beam with which the map was produced. 

Finally, we note that the region of the tip of the tail, where the North-South filament is clearly detected at 144 MHz and shows $\alpha_{\rm 54 MHz}^{\rm 144 MHz} \sim -1.2$, is masked with our 3$\sigma$ threshold. In fact, this structure is only barely detected at 380 MHz.

%###################################################

\section{Discussion}
\label{sec:discussion}

\begin{figure*}[t!]
\centering
    \includegraphics[scale=0.67]{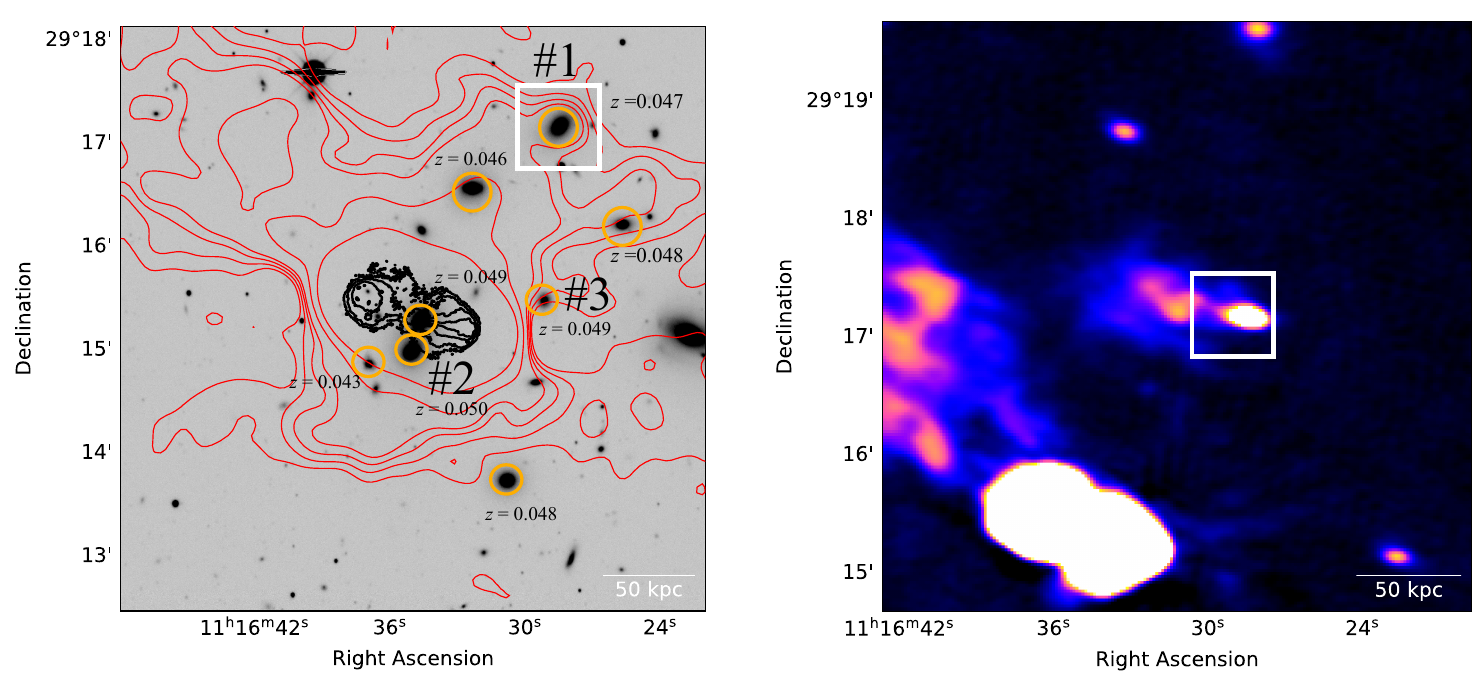}
    \caption{\textit{Left:} DR10 DESI Legacy Survey optical image of the inner region of A1213. Overlaid are the 54 MHz contours at $\sim 12''$ resolution (red) and the 144 MHz contours at $\sim$ 0.3$''$ resolution (black). Yellow circles denote members of A1213 that could be related to the elongated tail, with their corresponding redshift from SDSS. \textit{Right:} 144 MHz image at $\sim 6''$ resolution zoomed onto the elliptical galaxy on the North-West region of A1213. The optical galaxy is surrounded by a white square.}
\label{fig:optical}
\end{figure*}

In Fig. \ref{fig:large} we show the radio emission, as observed by LOFAR at 144 MHz, overlaid on the SDSS \textit{g}-band image. We have labelled relevant features such as the central AGN, the tail, the Eastern edge and the head-tail radio galaxy, which will be discussed in this section. 

\subsection{Origin of the tail}

To better visualise and investigate the interplay between thermal and non-thermal emission, in the left panel of Fig. \ref{fig:overlay} we show the LOFAR 54 MHz 3$\sigma$ contours overlaid on the adaptively smoothed XMM-\textit{Newton} map between 0.7 and 1.2 keV. The right panel shows a zoom-in on 4C\,29.41, where we have overlaid 144 MHz contours at 0.3$''$ resolution. As already reported in B23, the IGrM distribution is not typical of a cool-core system, where the morphology is usually roughly spherical and regular, with a clear core. Instead, we observe an elongated morphology and multiple emission peaks, with only few of them being associated with AGN counterparts. The thermal emission is stretched on the NE-SW direction. While there is no clear X-ray core, the X-ray centroid of the group as identified in B23 lies $\sim$30 kpc West to 4C\,29.41. The map apparently shows knotted X-ray emission surrounded by a lower surface brightness area. This is the result of the relatively short exposure time (i.e. low photon count), combined with the smoothing we have applied to enhance the thermal emission. For further details, we refer to B23.

The radio emission from the tail develops Eastwards from 4C\,29.41 for a total extent of $\sim$500 kpc, and apparently shows no clear interplay or spatial correspondence with features observed in the hot gas. The emission around 4C\,29.41 seems instead to be distributed following the major axis of the roughly elliptical gas distribution. This is in line with the findings reported in B23 and might suggests a sub-cluster merger in the NE-SW direction. Despite the presence of multiple X-ray peaks in the South-West part of the system, we detect no associated radio emission, either compact or extended, even though the morphology of the thermal emission closely resembles that observed in the North-East region, where non-thermal emission is present.

Previous studies on this system \citep[e.g.,][]{Giovannini_2009} first classified the tail as a radio halo. However, as also stated in B23, if this was the case it would be really peculiar, as it is off-centred and elongated with respect to the X-ray emission. Furthermore, it stands out from the X-ray luminosity - radio power correlation \citep{Giovannini_2009} usually observed for radio halos \citep{Feretti_2012, Cassano_2013, Cuciti_2021}. \citet{Hoang_2022} proposed instead that this emission could be the extended tail of the central radio galaxy. In this scenario, the tail could have reached its current extent thanks to the low IGrM density in that direction. They also suggest that the Eastern edge could have been generated by merger shocks interacting with the tail. While this interpretation sounds reasonable, given the spectral index map in Fig. \ref{fig:spidx}, B23 attributes the origin of the tail to a radio relic, i.e. the result of a shock that re-accelerated electrons in this region following first-order Fermi processes. They support this hypothesis through a spectral index map between a VLA 1.4 GHz observation \footnote{taken on March 6$^{\rm th}$, 2008 (C configuration) and June 2$^{\rm nd}$, 2008 (C/D configuration).} and the same LoTSS observation from which our data was derived. This map shows a gradient in the N-S direction, with the spectral index steepening from North to South, which would be evidence of a shock propagating in the same direction.

In both our spectral index maps (54-144 MHz and 144-380 MHz) we indeed detect hints of a gradient in the South-North direction, which can be noted from the zoom-in on the left panel of Fig. \ref{fig:spidx}. In this region, the spectral index is flatter in the Southern part of the tail, and it steepens moving North. This is opposite with respect to B23, which observed instead a flatter spectrum in the Northern region, which steepens moving South. Our higher-resolution spectral index and total intensity maps clearly show that the brightest emission of the tail is concentrated on the Southern region, where the spectral index is flatter. The less energetic electrons in the North are more likely the simple result of interaction with the surrounding medium and ageing due to radiative losses, rather than the consequence of shock re-acceleration, as instead suggested in B23. The distribution of the thermal gas strongly disagrees with the hypothesis of a radio relic, as its orientation is different from what we would expect from the X-ray morphology, that is stretched along the NE-SW axis: a shock would eventually re-accelerate electrons in a direction that would be almost perpendicular to that of the detected radio emission.

The 144 MHz 0.3$''$ map shows that the current outburst of 4C\,29.41 does not hint at any kind of physical connection with the tail. The LOFAR spectral index map shows hints of a spatial break between the AGN and the tail, which are separated by a small ($\sim$20 kpc width) region with very steep spectral index ($\alpha \sim$ -2.3). It is therefore likely that we are looking at two different sources, and that the tail did not directly develop from the central radio galaxy, or at least it does not originate from its current outburst. We suggest two possible scenarios for its origin.

\subsubsection{Non-thermal contribution from group members}

A first, possible explanation is that member galaxies near the position of 4C\,29.41 could have played a role for the formation of the tail. Indeed, we have inspected the Data Release 10 of the DESI Legacy Survey \citep{Dey_2019} at the position of A1213, and we found that there is a number of group member galaxies which could have possibly contributed to the process. In the left panel of Fig. \ref{fig:optical} we show the DESI DR10 image with overlaid radio contours, including LOFAR-VLBI. Redshifts are from spectra of SDSS DR16 \citep{Ahumada_2020}. An example is the source on the North-West with $z = 0.047$ (labelled \#1 in the figure) which, from SDSS, is an elliptical galaxy whose radio emission is visible even from 144 MHz high-resolution images (e.g. central panel of Fig. \ref{fig:radio}). If we push the HBA resolution (see right panel of Fig. \ref{fig:optical}), it is clearly visible that this source hosts an head-tail radio galaxy (as already discussed above) that, in the region of the radio core, looks physically disconnected from the extended emission on the East of the group. There are still hints of plasma mixing with the tail in the Eastern lobe, as also visible from the right panel of Fig. \ref{fig:optical}. Nevertheless, from the current data it seems unlikely that galaxy \#1 constitutes the major factor for the formation of the tail. Another contribution could have possibly come from an elliptical galaxy at $z = 0.05$ (labelled \#2 in the figure), which lies $\sim$20 kpc (in projection) South from 4C\,29.41. This source was already identified as a companion galaxy to 4C\,29.41 in B23. Our high-resolution image at 144 MHz shows that this galaxy does not exhibit radio emission on the same scale of 4C\,29.41.. Furthermore, the SDSS spectrum does not show any hint of the typical AGN emission lines, suggesting that the galaxy is not currently active, although it might have been in the past. Finally, it is worth noting that the source at \textit{z} = 0.049 West to 4C\,29.41 (labelled \#3 in the figure), identified as the BGG in B23, does not look to be connected with the tail, at least at low frequency. In summary, while radio emission from these galaxies (when present) could have possibly contributed to the formation of the tail, they hardly played the dominant role.

\subsubsection{Group weather and galaxy motions}

A second possible explanation is that the tail originated from a past AGN activity of 4C\,29.41, while the double-lobe emission represents the current outburst. This idea is supported by our spectral index maps, which show a sudden break in the gradient between 4C\,29.41 and the tail, as well as by the merger's orientation along the NE-SW axis, as suggested by the X-ray morphology (see B23). In this scenario, the active radio galaxy might be moving along this axis, leaving behind a trail of older electrons injected during a previous outburst. These features in A1213 closely resemble those observed in another dumb-bell galaxy, NGC 326, by \citet{Hardcastle_2019}. In that system, the Western wing appears confined, while the Eastern one has a tail-like extension with a filamentary structure and a sharp edge, similar to what we see in A1213. They propose that large-scale hydrodynamical processes, possibly combined with black holes interaction, could explain the observed morphology. In A1213, an initial outburst of 4C\,29.41 could have released a population of seed electrons, while the galaxy continued moving southwest along the line of sight (B23). Following this first event, a second, more recent outburst likely produced the double-lobed structure resolved at 0.3$''$ resolution. If this is the case, the tail might consists of old, re-energised AGN plasma, akin to radio phoenixes sometimes found in disturbed galaxy clusters \citep[e.g.,][]{Ensslin_2002, deGasperin_2015, Mandal_2019, Pasini_2022a}. The observed filaments may trace magnetic field enhancements possibly caused by the ongoing merger. Additionally, the orientation of the Eastern edge suggests thermal pressure is preventing further expansion of the tail, bending the structure. In conclusion, the current environment of A1213 likely results from a combination of multiple AGN outbursts and group dynamics, with mergers and gas motions shaping the non-thermal emission.

While it is not trivial to conclusively prove this scenario, we can perform a number of tests to check and confirm that the tail is not physically related to the current outburst of 4C\,29.41. First of all, to eventually trace gradients in the spectral shape of the electron energy distribution between lower and higher frequency, we have produced a spectral curvature map (SC map), similarly to e.g. \citet{Rajpurohit_2021}, by deriving the curvature in each pixel as:

\begin{equation}
{\rm SC} = -\alpha_{\rm low} + \alpha_{\rm high}
\end{equation}

\begin{figure}[h!]
\centering
    \includegraphics[scale=0.37]{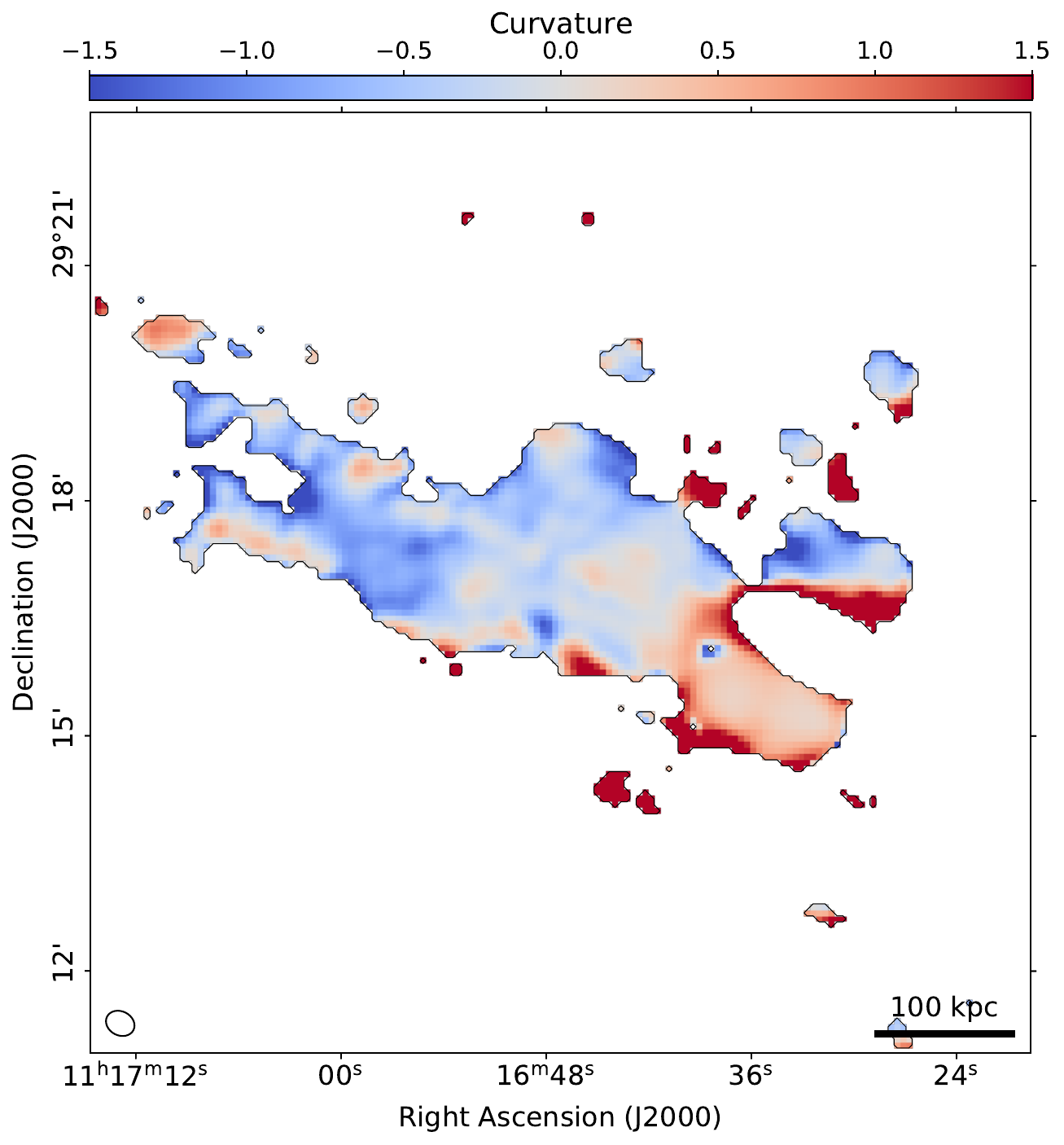}
    \caption{Spectral curvature (SC) map between 54-144 MHz and 144-380 MHz at $\sim 20''$ resolution. A negative value implies a convex spectral shape.}
\label{fig:curvature}
\end{figure}

This was applied to spectral index maps between 54-144 MHz and 144-380 MHz by convolving them at the smallest common beam where the extended emission is clearly visible, i.e. with $\sim 20'' \times 20''$ resolution. The SC map is shown in Fig. \ref{fig:curvature}.

\begin{figure*}[t!]
\centering
    \includegraphics[scale=0.35]{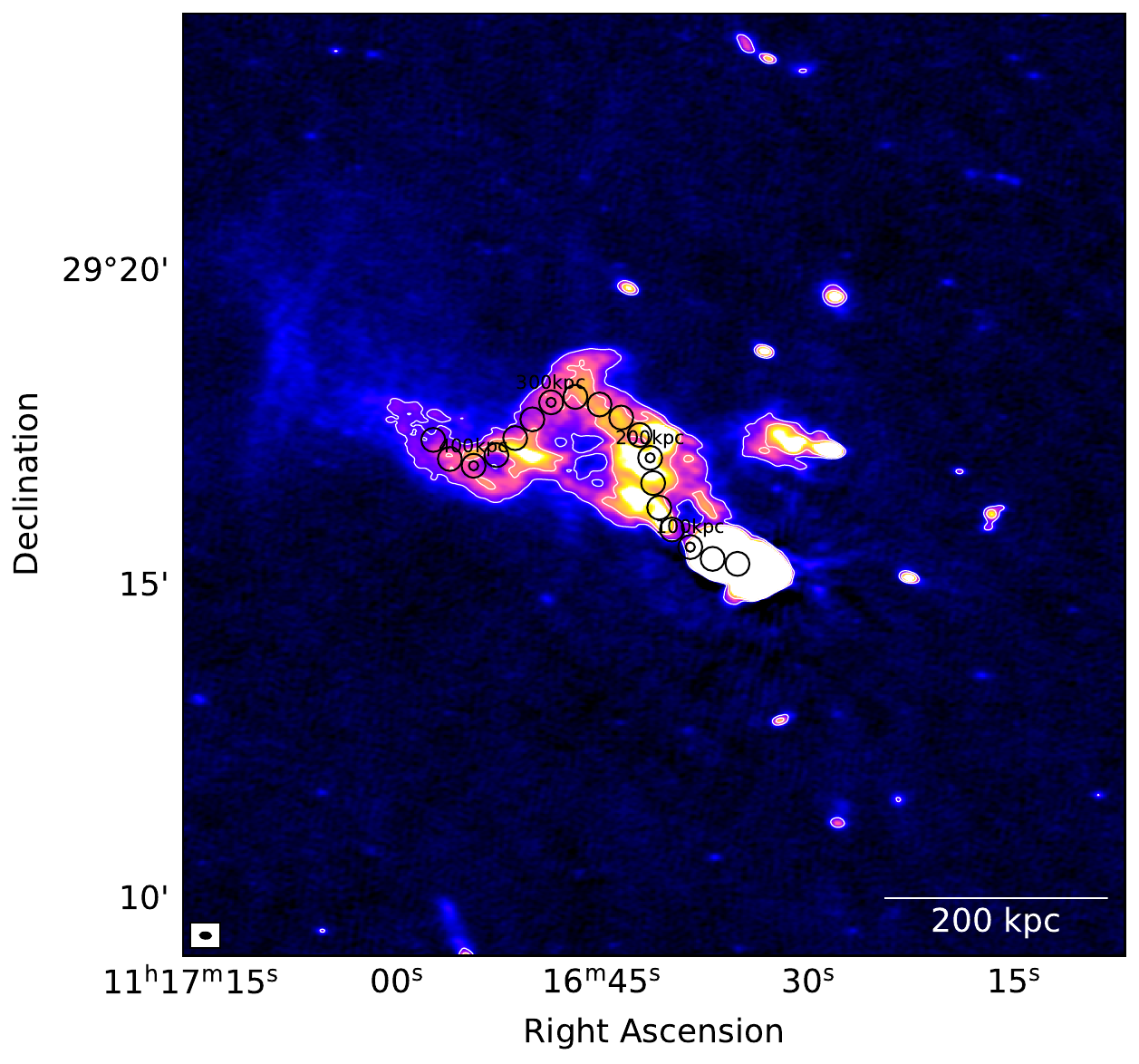}
    \hspace{1cm}
    \includegraphics[scale=0.6]{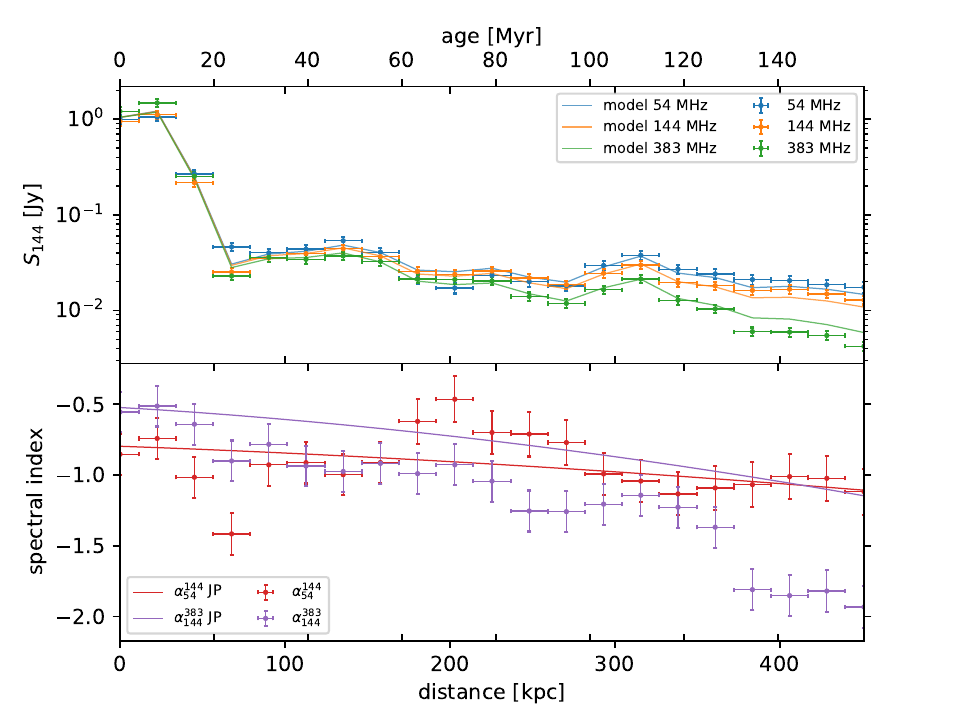}
    \caption{\textit{Left}: The image shows the regions from which the spectral index was sampled and fit with a JP model, overlaid on the 144 MHz map. \textit{Right}: The top panel shows the flux density at different frequencies for each region, while the bottom panel shows the corresponding spectral index and the fit with a JP model.}
\label{fig:sampling}
\end{figure*}

The "edges" of the radio emission are affected by large errors from the spectral index maps. For this reason, we will not comment on the significantly higher (or lower) SC values in the regions surrounding the black contours. In the region of 4C\,29.41 the SC values are mostly positive or close to 0, implying that the spectrum is not significantly steepening from low to high frequency, as expected from a situation where the AGN outburst is most likely still occurring (or has occurred recently). On the other hand, the tail shows hints of a negative curvature, which becomes more significant as we move Eastwards: the high-frequency spectral index is becoming steeper and steeper, compared to the low-frequency index. Most importantly, we also note a rather sharp break in the SC distribution between the tail and the region of 4C\,29.41, as the SC suddenly decreases from $\sim$0.5 to $\sim$ -0.3. This supports the scenario where the double-lobed structure is physically disconnected from the elongated emission, as we are detecting two different outbursts that happened on different timescales.

Another hint that the two emissions are currently unrelated, and in fact might come from different outbursts of the same central engine source, comes from the synchrotron ageing models. In the absence of particle re-acceleration, the electron population is expected to age due to radiative (synchrotron and Inverse Compton, IC) and adiabatic losses. In this "pure-aging" scenario, if the tail originated from the current outburst of 4C\,29.41, we should observe a spatial evolution of the spectral index along the extended emission moving Eastwards from the AGN. In more detail, the spectral index should be flatter ($\alpha \sim -0.6$) in the injection point and then gradually become steeper as we move along the tail. To test this hypothesis, we have sampled the flux density starting from the injection point of the electrons, which we assumed spatially coincident with the AGN core in the case where the tail directly develops from 4C\,29.41, and moving along the extended emission, as shown in the left panel of Fig. \ref{fig:sampling}. Each sampling region is as large as the smallest common beam ($23'' \times 23''$ ) of 54, 144 and 380 MHz images, which were suitably produced with a matched \textit{uv-}cut as already discussed in Sec. \ref{sec:spindex}. For each region, we estimate the spectral index between 54 and 144 MHz and between 144 and 380 MHz. The spectral index distribution is then fitted through a Jaffe-Perola (JP) model \citep{Jaffe_1973}, for which we assumed that the break frequency is the same at a given distance. In this model, as we assume an injection index $\alpha = -0.8$ from our spectral index map, the ageing of the electrons population is dependent only upon the magnetic field strength and the projected velocity of the host galaxy through the cluster environment, i.e. the IGrM. To maximise the lifetime of the electron population (so that we get is an upper limit on the radiative age), we have assumed a minimum loss magnetic field $B_{\rm min} = B_{\rm CMB} / \sqrt{3}$, with $B_{\rm CMB} = 3.2 \times (1+z)^2 \ \mu G$ being the magnetic field strength of the Cosmic Microwave Background (CMB) at redshift \textit{z}. We get $B_{\rm min} = 2.03 \ \mu G$, which is consistent with the estimate of 2-3 $\mu G$ reported in B23, and at this point the only free parameter is the galaxy velocity (see also \citealt{Edler_2022}). The result is shown in the right panel of Fig. \ref{fig:sampling}.

The synchrotron ageing model clearly fails to reproduce the observed behaviour of the spectral index. We obtain in fact a galaxy velocity of $\sim$2785 km/s, and a reduced $\chi ^2$ -squared of $\chi ^2 /$DoF = 1641 (where DoF are the degrees of freedom), which is an obvious indicator that the model is not providing a correct description of the data. The flux density distribution along the extended emission follows indeed a non-trivial behaviour, exhibiting a sharp and sudden decrease as soon as we move out of the brightest region, which we would not expect if the tail directly originated from it. The spectral index distribution is also complex: the current outburst shows a convex spectrum (as also found in Fig. \ref{fig:curvature}), with the higher-frequency spectral index being flatter than the low-frequency one\footnote{Although errors remain quite large.}. On the other hand, moving along the tail we find a more typical trend, with a gradual steepening. However, we also note that the outermost part of the tail ($\sim$300 kpc onwards from the AGN core) shows a rather constant spectral index at low-frequency, while at higher frequency it exhibits the typical steepening of the synchrotron spectrum, and it reaches a plateau at higher distance from the (supposedly) injection point. While similar cases of tails mildly re-energised by turbulence have already been found \citep{deGasperin_2017}, the complex spectral index distribution observed for this source prevents us from drawing further conclusions.

Nevertheless, it is clear that a simple pure-aging model along the AGN and tail, assuming that the latter originated from the same outburst that is currently supporting the former, is not able to reliably describe the spectral shape. The most obvious explanation for the observed distribution is that we are indeed looking at two different sources.

\subsection{The synchrotron spectrum of 4C\,29.41}
\label{sec:synch}

\begin{table}
\scriptsize
\caption{Integrated flux density and spectral indices of 4C\,29.41 at 54, 144 and 380 MHz.}
\begin{tabular}{llllll}
\hline\hline
&$S_{54 \rm MHz}$&$S_{144 \rm MHz}$&$S_{380 \rm MHz}$&$\alpha_{54 \rm MHz}^{144 \rm MHz}$&$\alpha_{144 \rm MHz}^{380 \rm MHz}$\\
& [Jy] & [Jy] & Jy & & \\
\hline
\hline
No \textit{uv}-cut & 15.5 $\pm$ 1.6 & 6.6 $\pm$ 0.7 & 3.8 $\pm$ 0.2 & -0.87 $\pm$ 0.15 & -0.57 $\pm$ 0.12\\
\textit{uv}-cut & 13.0 $\pm$ 1.3 & 6.4 $\pm$ 0.7 & 3.6 $\pm$ 0.2 & -0.72 $\pm$ 0.15 & -0.59 $\pm$ 0.13\\
\hline
\end{tabular}
\tablefoot{The first three columns show the 54, 144 and 380 MHz integrated flux density of 4C\,29.41, without and with applying a \textit{uv}-cut. The last two columns show the corresponding spectral indices for both cases.\\}
\label{tab:fluxes}
\end{table}

We have measured the integrated flux density within 3$\sigma$ contours of 4C\,29.41 at all our three frequencies\footnote{At the same resolution.} to study in more detail its synchrotron spectrum. The results are summarised in Table \ref{tab:fluxes}.

If we estimate the spectral index using these values, we find $\alpha_{54 \rm MHz}^{144 \rm MHz} = -0.87 \pm 0.15$ and $\alpha_{144 \rm MHz}^{380 \rm MHz} = -0.57 \pm 0.12$, which confirms the positive curvature already found in Fig. \ref{fig:curvature}. Furthermore, we also retrieved the 1.4 GHz image from the FIRST survey (Faint Images of the Radio Sky at Twenty-Centimeters, \citealt{Becker_1994}) and estimated the high-frequency spectral index between 380 MHz and 1.4 GHz, $\alpha = -0.53 \pm 0.09$, which further confirms the convex shape. Given the morphology and extension of the radio galaxy, a flattening of the spectrum moving to higher frequency is somehow unexpected.

First, we checked the flux scale of our images, either by using available surveys at close frequencies (e.g. the TIFR GMRT Sky Survey at 150 MHz, TGSS, \citealt{Intema_2017}) or by re-scaling the flux density of point sources in the field assuming a typical spectral index $\alpha = -0.8$, and confirmed that our data do not show any obvious flux scale mismatch exceeding the expected systematic uncertainties (see Sec. \ref{sec:calib}).

\begin{figure}
\centering
    \includegraphics[width=24em, height=24em]{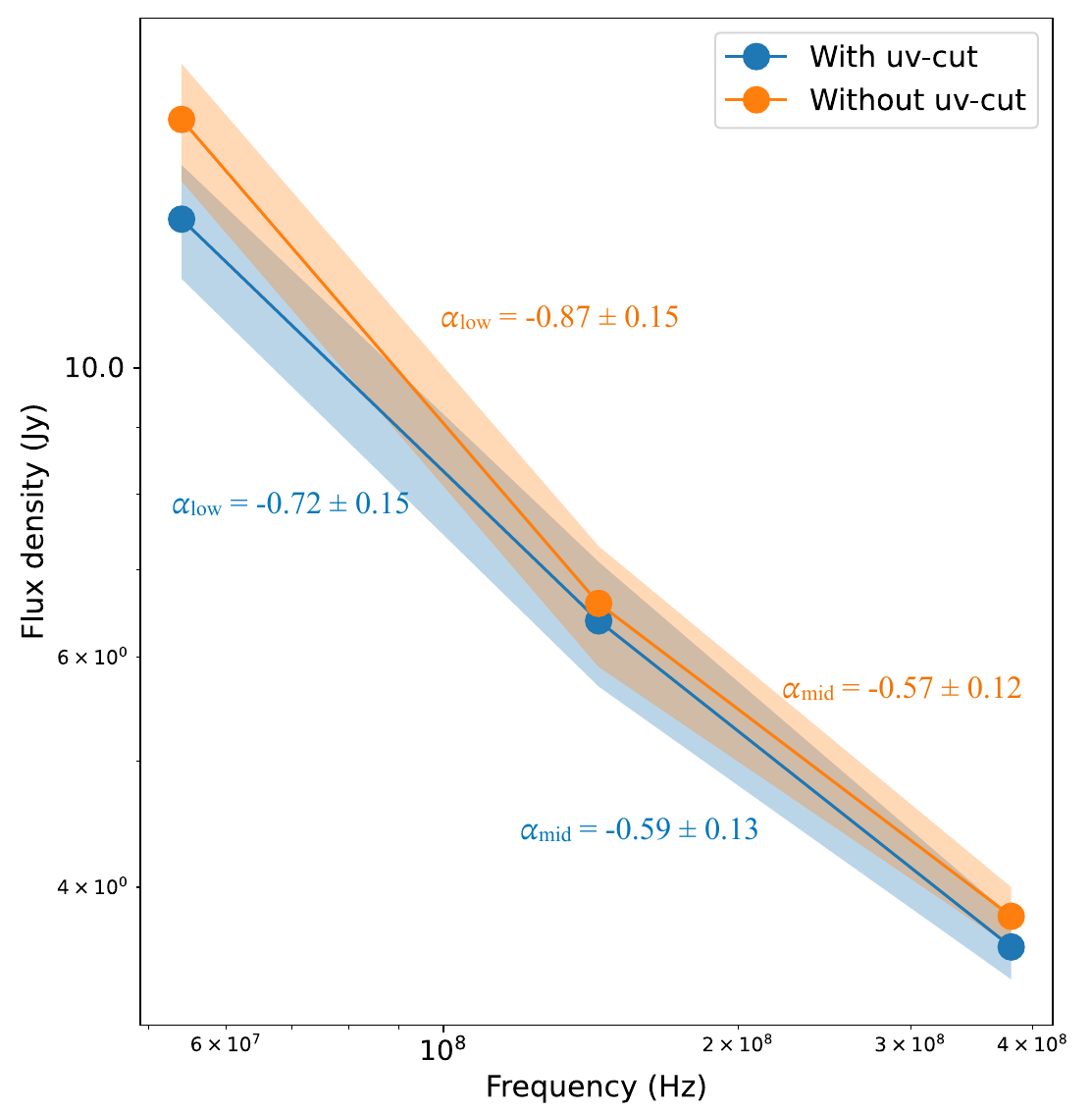}
    \caption{Synchrotron spectrum of 4C\,29.41 between 54 and 380 MHz. The orange curve shows the case where no \textit{uv}-cut was applied while imaging, thus including the contribution of extended emission, that increases the flux density especially at lower frequency. The blue curve shows instead the case with a minimum \textit{uv}-cut of 1.6k$\lambda$, which roughly corresponds to 2$'$ (i.e. larger emission is not present).}
\label{fig:spectrum}
\end{figure}

We then investigated whether the observed spectrum could be the result of the combination of two components, one steeper and one flatter. The companion galaxy of 4C\,29.41 is slightly offset from the radio galaxy; furthermore, if we separately measure the spectral index of the two lobes, we find that the two spectra are consistent with each other and the observed flattening at higher frequency exists for both lobes. We have then considered the hypothesis that extended emission from the tail might be contaminating the region of the radio galaxy (i.e. adding spurious flux). In this scenario, the lower frequencies should be more affected by this issue, because of the spectral shape of the tail. We should be able to remove most of the extended sources contribution by applying a specific \textit{uv}-cut when cleaning, so that sources above a certain LLS are not imaged. Since the LLS of the radio galaxy is $\sim$2$'$, we have produced images at 54, 144 and 380 MHz by cutting all visibilities below 1.6k$\lambda$, and measured again the integrated flux density. In this way, only the radio galaxy is imaged, and we lose larger emission. Results are listed in Table \ref{tab:fluxes} and shown in Fig. \ref{fig:spectrum}.

The 54 MHz data is particularly affected by this problem, since the extended emission primarily shines at lower frequency and is clearly detected, significantly contaminating the flux density measurement of the radio galaxy. On the other hand, we find a less dominant effect at 144 and 380 MHz. By removing the contribution of the large scale emission from the tail, the synchrotron spectrum of the radio galaxy in the the range 54-300 MHz is consistent with a power-law with $\alpha \sim -0.7$ within uncertainties, which is typical for this kind of sources. Finally, it is worth mentioning that the flattening observed when accounting for the FIRST measurement at 1.4 GHz is instead probably driven by the core, which starts to dominate the total source flux density at these higher frequencies.

\subsection{The Eastern edge: tip of the tail or radio relic?}
\label{sec:relic}

\begin{figure*}[t!]
\centering
    \raisebox{0.15cm}{\includegraphics[scale=0.38]{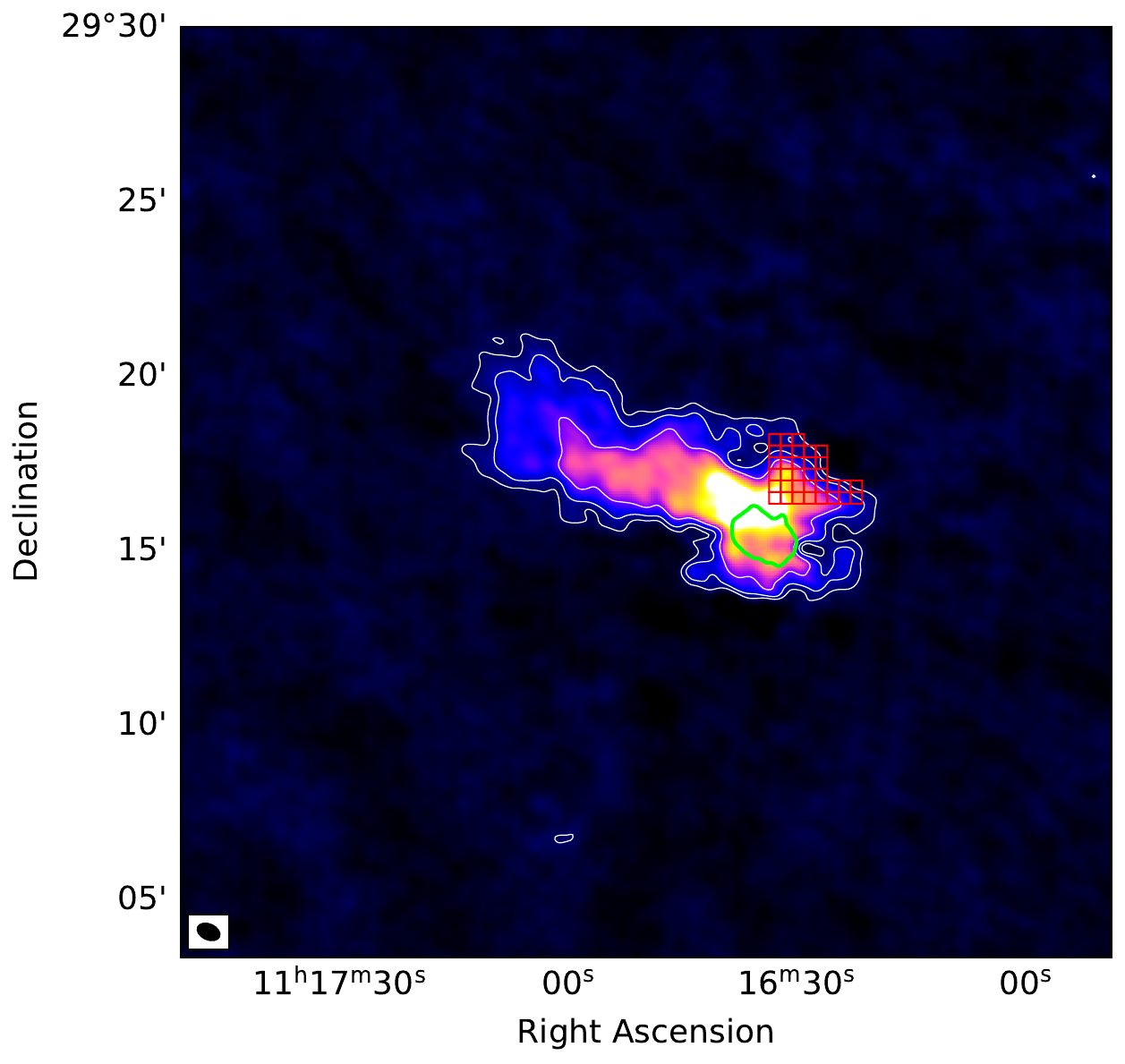}}
    \hspace{1cm}
    \includegraphics[scale=0.34]{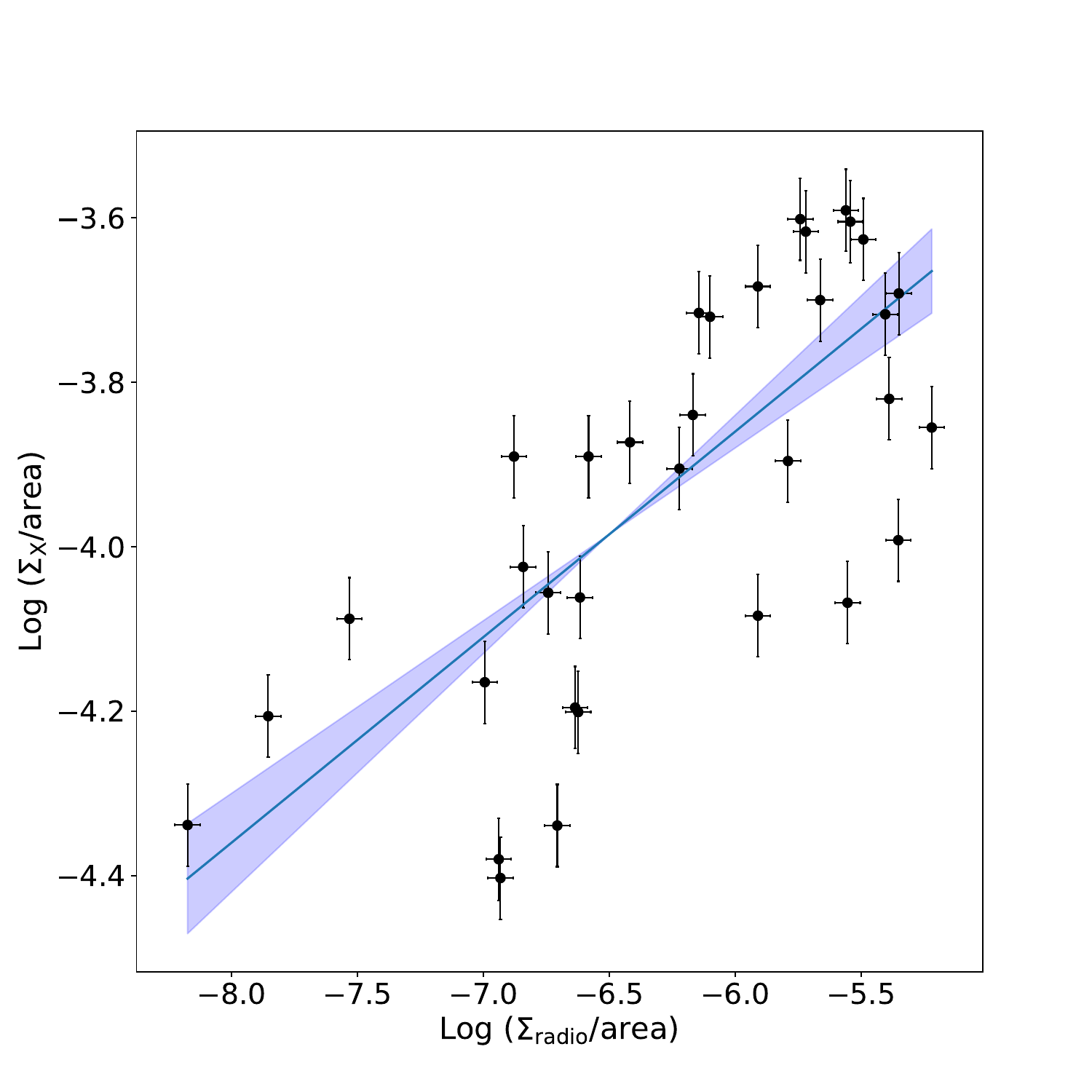}
    \caption{\textit{Left:} Source-subtracted map of A1213 at 54 MHz, where all sources with LLS $<$ 250 kpc have been removed. It is not possible to remove the whole emission from 4C\,29.41, which blends together with the tail and with the roundish emission on the Western region. The red grid shows the region of interest of the point-to-point analysis, while the green contours encircle the region where 4C\,29.41 is located. \textit{Right:} Result of the point-to-point analysis between the X-ray and the LOFAR LBA source-subtracted image of A1213. The surface brightness is estimated as the sum of every pixel in each square of the grid, divided by its area. The blue line shows the best fit in the form $Y = kX + A$, where $k = 0.25 \pm 0.04$ and $A = -2.36 \pm 0.26$.}
\label{fig:ptrex}
\end{figure*}

The North-South oriented edge detected at 144 MHz in the Eastern part of the tail shows the typical morphology of a radio relic. Its orientation with respect to the IGrM (and therefore to the merger axis) is also in agreement with this hypothesis. Indeed, \citet{Hoang_2022} first suggested the presence of a shock in the East-West direction, which could have generated the source. However, the region is too thin and our resolution not high enough to detect any kind of spectral index gradient in the LOFAR map. We have therefore estimated the integrated flux density within 3$\sigma$ contours at 54 and 144 MHz, finding 513 $\pm$ 51 mJy and 126 $\pm$ 13 mJy, respectively, which yields $\alpha = -1.4 \pm 0.15$. While this is obviously in agreement with the spectral index map in the left panel of Fig. \ref{fig:spidx}, it is only marginally consistent with the usual integrated values ($\sim -1.2$) observed in radio relics (see e.g. \citealt{vanWeeren_2019}).

This structure is not detected by uGMRT at 380 MHz. Only by performing a significant tapering of the visibilities we start to observe faint hints of the source, although image negatives affect this region. Therefore, it is not possible to accurately estimate the integrated flux density at this frequency. Nevertheless, we can assume that the spectral index is the same than what found at lower frequency, and predict the expected flux density value: this yields an integrated flux density of $\sim$32 mJy at 380 MHz. Given the \textit{rms}-noise of the uGMRT image\footnote{$\sim$0.3 mJy beam$^{-1}$ if we taper at 30$''$.}, we should clearly be able to detect it if this was the case, which suggests that the real value is lower and, therefore, the spectral index is steeper. Indeed, if we use the same 3$\sigma$ contours exploited at 144 MHz and the \textit{rms} noise mentioned above, we get an upper limit on the integrated flux density of $\sim$25 mJy, which consequently leads to an upper limit on the spectral index $\alpha_{144 \rm MHz}^{380 \rm MHz} < -2.1$. This steepening is not commonly observed in radio relics. Furthermore, no X-ray discontinuity was identified in B23 which could hint at the presence of a shock propagating through this region. We can conclude that this source is just the final part of the tail. The different orientation is likely due to the interaction with the surrounding IGrM, which has bent the structure. This also explain the steeper spectral index observed in Fig. \ref{fig:spidx}.

\subsection{Does A1213 host diffuse radio emission?}

B23 reports the presence of fragmented radio emission close to the group X-ray centroid, around the BGG, which they attribute to a candidate radio halo, although their data did not allow to investigate further. Their 144 image is affected by calibration artefacts in this region, most likely because of the brightness of 4C\,29.41. On the other hand, our calibration strategy significantly improved this aspect. Once artefacts are removed, the BGG shows a rather compact radio emission, with no hints of diffuse structures.

However, we note that the left panel of Fig. \ref{fig:radio} shows the presence of roundish (although slightly elongated on the N-S axis) diffuse radio emission around 4C\,29.41, which is visible even at high resolution, and especially from the source-subtracted map shown in the left panel of Fig. \ref{fig:ptrex}. The same emission is also detected in the LOFAR-VLBI image in the left panel of Fig. \ref{fig:VLBI}. While it is possible that it could be part of the inner side of the tail, the different morphology might also hint at a different origin. Unfortunately, we had to mask this region when producing the spectral index maps in Fig. \ref{fig:spidx}, because of the presence of calibration artefacts and negatives in the 144 and 380 MHz image. Therefore, it is currently not possible to constrain the synchrotron spectrum of the diffuse emission, nor to put upper limits. Nevertheless, we can try to speculate on its nature based on the currently available data.

A possibility is that this emission first originated from 4C\,29.41, and then got trapped in the central region of the group, evolving to a low surface brightness source that might resemble clusters' mini-halos \citep{vanWeeren_2019}. In this scenario, seed fossil electrons are provided by the central AGN. It is also possible that other active group members might have contributed in the past, although SDSS spectra of galaxies in Fig. \ref{fig:optical} do not currently show any emission line typical of AGN.

A well-known physical correlation, which directly arises from the connection between mini-halos and ICM motions, exists between their radio power and the host cluster X-ray luminosity (see e.g. \citealt{Biava_2021, Riseley_2022}). We have therefore exploited {\ttfamily PT-REX} \citep{Ignesti_2022} to trace the radio and X-ray surface brightness of the diffuse emission through a point-to-point analysis \citep{Govoni_2001, Rajpurohit_2021}. To avoid as much as possible calibration artefacts and contamination from 4C\,29.41, the algorithm was applied to the 54 MHz source-subtracted data, which is the cleanest in this region, and leftover spurious emission in the centre of the AGN and in the region of the head-tail in the North was masked. The algorithm designs a grid above the region of interest, which is shown in the left panel of Fig. \ref{fig:ptrex}, and sample the surface brightness in each square of the grid from the radio and X-ray images. The grid was carefully selected to avoid the region of 4C\,29.41 and the Southern part of its radio emission, i.e. the proximity of the location of the companion galaxy before the source subtraction. We put instead the focus on the Northern region where the AGN should contaminate less. The result is shown in the right panel of Fig. \ref{fig:ptrex}.

The surface brightness values show a correlation, albeit scattered. The best fit, which was computed in the form $Y = kX + A$, shows $k = 0.25 \pm 0.04$ and $A = -2.36 \pm 0.26$, i.e. a sub-linear correlation which is usually typical of giant halos, rather than mini-halos. The Pearson and Spearman indices are both $\sim$0.72, which translates into a very low \textit{p}-value of $\sim 10^{-5}$. While this might indicate a real correlation, it is also worth noting that the analysis had to be restricted to a very small region, because of possible AGN contamination. For the same reason, we are not sampling the Southern part of the emission. Therefore, it is currently hard to derive strong conclusions about its origin. While the hypothesis of a mini-halo in a galaxy group is tempting, especially considering the observed physical and spatial correlation with the thermal gas, it is not possible to exclude the possibility that this emission might just constitute the inner part of the tail, or that it might just be plasma being transported by the "weather" of the IGrM, similarly to e.g. NGC 507 \citep{Brienza_2022}. Finally, it is also worth noting that mini-halos are usually observed in systems hosting cool cores \citep[e.g.,][]{Govoni_2009, Biava_2021}, whereas A1213 shows multiple X-ray peaks and no clear core.

%###################################################

\section{Conclusions}

In this work, we have investigated the low-frequency radio emission in the galaxy group A1213 by exploiting proprietary LOFAR 54 MHz and uGMRT 380 MHz observations. As found from previous studies (e.g. B23), this system is a  galaxy group with $\sim$143 galaxy members and a disturbed and elongated X-ray morphology. We have complemented our data with reprocessed LOFAR 144 MHz observations at both 6$''$ and 0.3$''$ resolution from LoTSS, and with an archival XMM-Newton observation. Our results can be summarised as follows:

\begin{itemize}

\item A1213 exhibits complex radio emission at low frequency. One of the brightest group members, 4C\,29.41, which is classified as a dumb-bell galaxy (i.e. it has two optical nuclei), hosts a bright FRII radio galaxy with two symmetric lobes. Only the Western jet is detected from our 144 MHz 0.3$''$ images, likely because of relativistic boosting. From the same region of 4C\,29.41, a striking $\sim$500 kpc-long trail of emission extends Eastwards. 

\item The trail is detected at all our three frequencies and shows multiple emission peaks along its length. The spectral index maps reveal that the spectral index steepens from the AGN core down to the edge of the emission, reaching $\alpha \sim$ -2.5 in the outer side and showing a sudden break between the region of 4C\,29.41 and the tail. A roundish spot of flatter emission is detected on the tip. Our evidence support the hypothesis that the source is a radio tail which directly extends from the region of 4C\,29.41, rather than a radio relic as previously suggested in recent studies. 

\item We have produced a spectral curvature (SC) map and studied the spectral index distribution along the tail by fitting it with a standard JP synchrotron model. Together with the well-confined morphology of the FRII radio galaxy, this evidence suggests that the tail did not directly originate from the current outburst of 4C\,29.41, but it is likely the result of a past activity of its central engine, which expanded along a less dense region of the IGrM. On a later phase, the galaxy kept moving along the line of sight, producing the current FRII morphology in a more recent outburst. On the other hand, the old plasma was left behind and started to shine again due to re-acceleration mechanisms, as suggested from the spectral index plateau observed in Fig. \ref{fig:sampling}. The current state of A1213 might therefore be the result of a combination of galaxy motions, re-started AGN activity and group weather.

\item The thermal emission from the IGrM shows multiple peaks of emission and an elongated SW-NE morphology, with no obvious core. The X-ray centroid was previously identified in B23 to be $\sim$30 kpc West of 4C\,29.41. The non-thermal emission is somehow present only in the North-East region of the hot IGrM, while in the South-West of the group no radio source is detected at all frequencies.

\item In the Eastern part of the tail we observe a thin ($\sim$30 kpc) filament oriented in the North-South direction. This structure is not detected at 380 MHz unless by performing a significant tapering of the visibilities. Its location and morphology might be indicative of a radio relic. We find an integrated spectral index $\alpha \sim -1.4$ between 54 and 144 MHz which, however, significantly steepens at higher frequency, which is not typical of relics. This, together with the physical connection observed with the tail, likely excludes the hypothesis of a diffuse source, and points instead to an interaction of the terminal part of the tail with the surrounding IGrM.

\item We investigated the nature of the emission around 4C\,29.41, which was already tentatively detected in B23 with 144 MHz data. Candidate diffuse emission is clearly observed at 54 MHz and 144 MHz (although only from 1.5$''$ resolution images) thanks to our calibration strategy. The point-to-point analysis between the radio and the X-ray surface brightness seems to hint at a possible physical connection between the thermal gas and non-thermal plasma. In this scenario, a plausible hypothesis would be that of a mini-halo, whose seed electrons might have been injected by the activity 4C\,29.41 itself. Nevertheless, AGN contamination, even after source subtraction, prevents us from performing a thorough spectral analysis, which would help to assess the true nature of this emission. The hypothesis that this source could actually be just the inner part of the tail, or plasma transported by the IGrM, cannot currently be excluded. 

\end{itemize}

A1213 is only the first galaxy group, among the X-GAP sample, that we investigate at low radio frequency thanks to the available wealth of data. However, it already revealed a plethora of interesting features. Our study confirms, once more, that galaxy groups are not merely a scaled-down version of galaxy clusters, and that it is essential to focus on the interplay between thermal and non-thermal emission even in this lower-mass regime. Combining low-frequency radio data with X-ray observations can provide crucial insights into the role of AGN in the evolution of galaxy groups, and help to assess the outburst history that has shaped their current state.

% acknowledgements

\begin{acknowledgements}

We thank the referee for useful comments and suggestions. The LOFAR LBA data utilized in this study was awarded as a prize for winning the 2022 LOFAR Boat Race held in Cologne. FdG acknowledges the support of the ERC Consolidator Grant ULU 101086378. DH is supported by the Deutsche Forschungsgemeinschaft (DFG, German Research Foundation) under research unit FOR 5195: “Relativistic Jets in Active Galaxies”). RS acknowledges the support of the Department of Atomic Energy, Government of India, under project no. 12-R\&D-TFR-5.02-0700. The research leading to these results has received funding from the European Union’s Horizon 2020 research and innovation programme under grant agreement No 101004719 [ORP].

% LOFAR
The Low Frequency Array, designed and constructed by ASTRON, has facilities in several countries, that are owned by various parties (each with their own funding sources), and that are collectively operated by the International LOFAR Telescope (ILT) foundation under a joint scientific policy.

% GMRT
We thank the staff of the GMRT that made these observations possible. GMRT is run by the National Centre for Radio Astrophysics of the Tata Institute of Fundamental Research.

% ADS
This research has made use of NASA's Astrophysics Data System.

% DS9
This research has made use of SAOImage DS9, developed by Smithsonian Astrophysical Observatory.

\end{acknowledgements}

\bibliographystyle{aa.bst}
\bibliography{bibliography}

\begin{appendix}
\normalsize

\section{Spectral index error maps}
\label{app:a}

We show here the spectral index uncertainty maps corresponding to Fig. \ref{fig:spidx}. The spectral index and corresponding errors were estimated using:

\begin{equation}
\label{eq:spindex}
\alpha_{\nu1}^{\nu2} = \dfrac{\ln S_1 - \ln S_2}{\ln \nu_1 - \ln \nu_2} \pm \dfrac{1}{\ln \nu_1 - \ln \nu_2} \sqrt{\bigg(\frac{\sigma_1}{S_1}\bigg)^2 + \bigg(\frac{\sigma_2}{S_2}\bigg)^2} ,
\end{equation}

where $S_1$ and $S_2$ are the flux densities at frequencies $\nu_1$ and $\nu_2$, respectively, while $\sigma$ is the corresponding error including flux scale uncertainties (see Sec. \ref{sec:calib}).

\begin{figure}[h!]
\centering
    \includegraphics[scale=0.35]{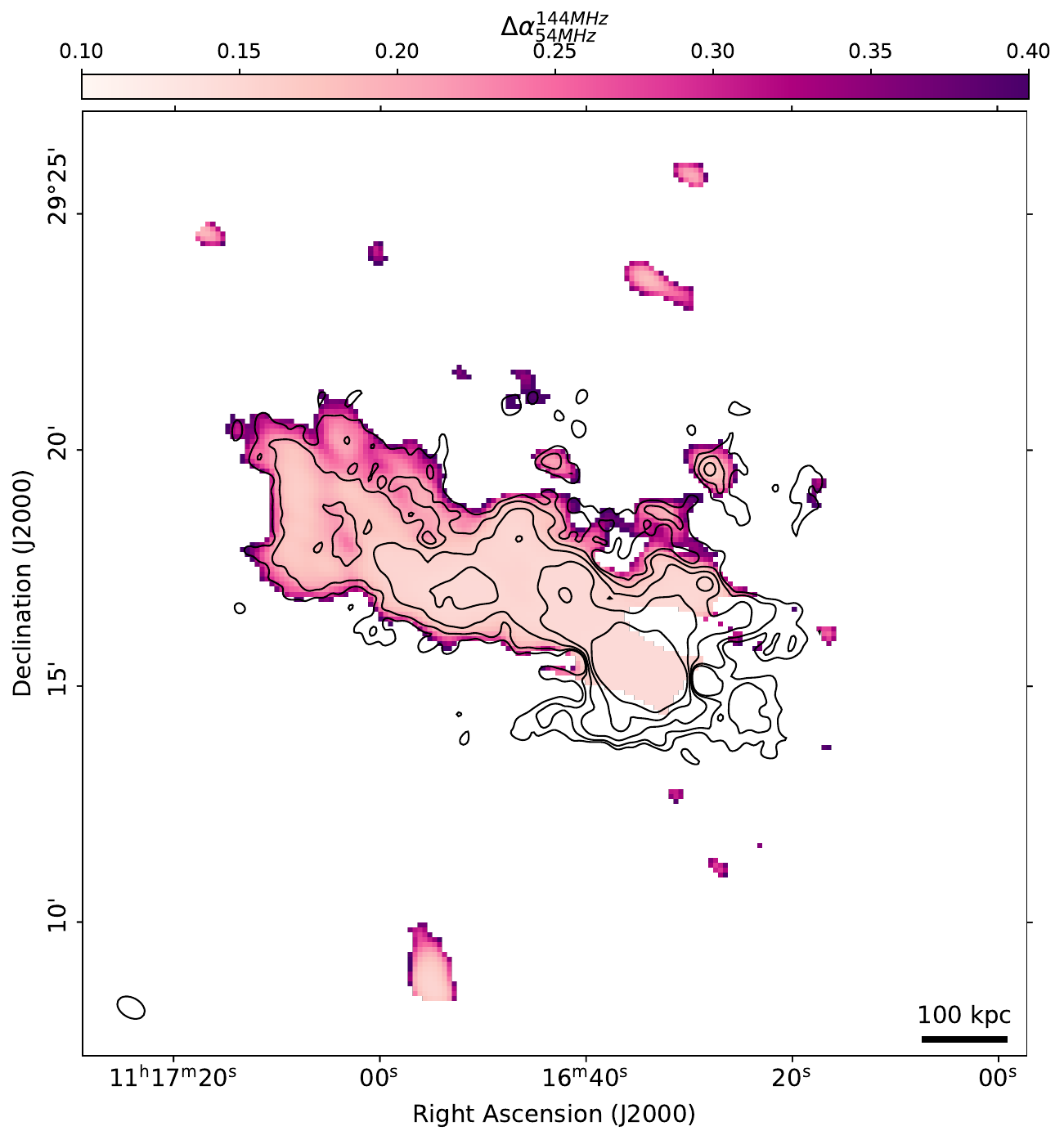}
    \includegraphics[scale=0.35]{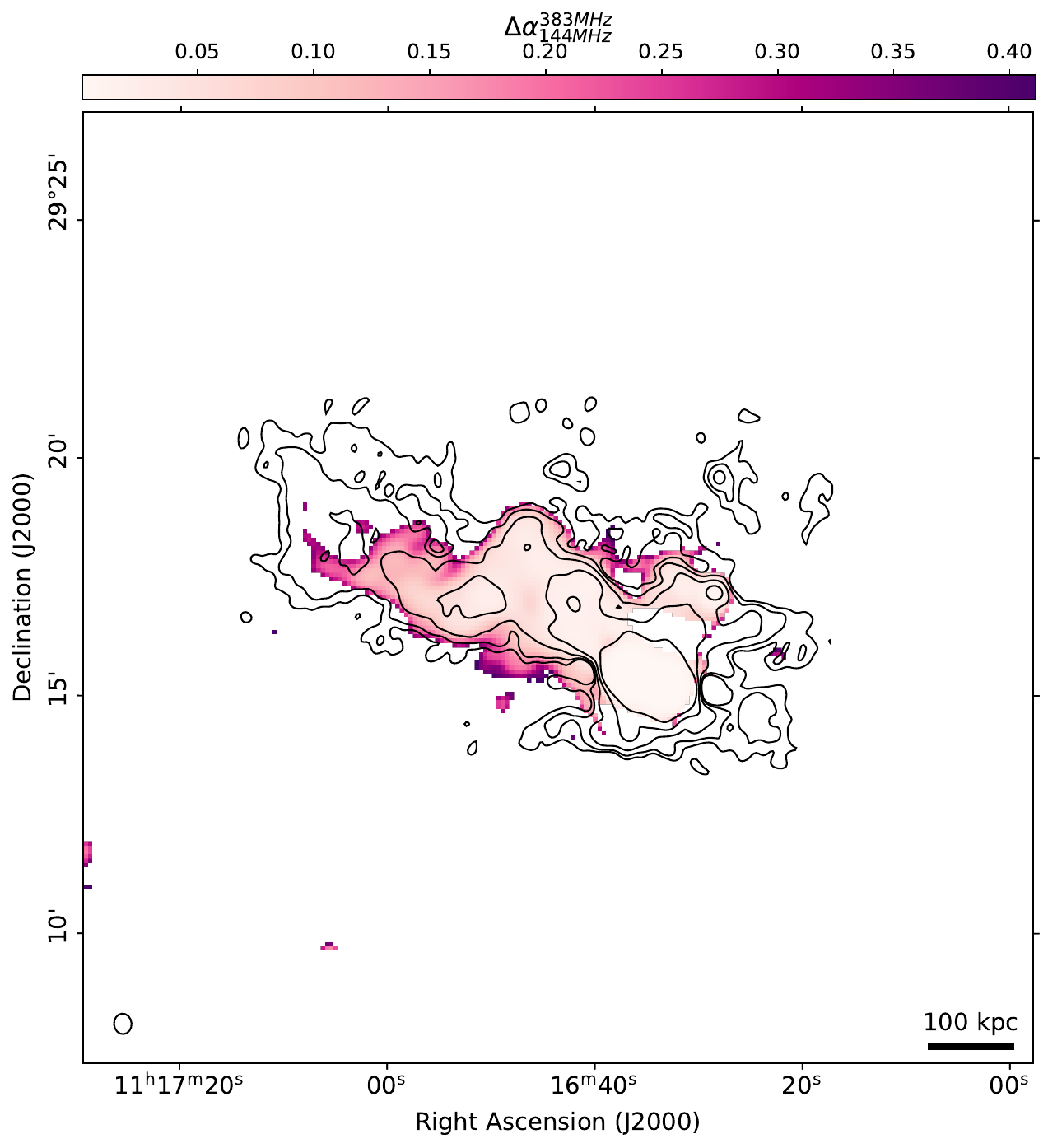}
    \caption{\textit{Top:} Spectral index error map between 54 and 144 MHz. Overlaid in black are 54 MHz contours. \textit{Bottom:} Spectral index error map between 144 and 380 MHz. Overlaid in black are 54 MHz contours.}
    \label{fig:appendix}
\end{figure}

\end{appendix}

\end{document}